\documentclass[english,aps,manuscript,reprint]{revtex4-1}
\usepackage[T1]{fontenc}
\usepackage[latin9]{inputenc}
\usepackage{amsbsy}
\usepackage{amstext}
\usepackage{graphicx}
\usepackage{esint}

\makeatletter

\@ifundefined{textcolor}{}
{%
 \definecolor{BLACK}{gray}{0}
 \definecolor{WHITE}{gray}{1}
 \definecolor{RED}{rgb}{1,0,0}
 \definecolor{GREEN}{rgb}{0,1,0}
 \definecolor{BLUE}{rgb}{0,0,1}
 \definecolor{CYAN}{cmyk}{1,0,0,0}
 \definecolor{MAGENTA}{cmyk}{0,1,0,0}
 \definecolor{YELLOW}{cmyk}{0,0,1,0}
}

\providecommand \href@noop [0]{\@secondoftwo}%

\makeatother

\usepackage{babel}
\begin{document}

\title{Stochastic entropy production arising from nonstationary thermal
transport}

\author{Ian J. Ford, Zachary P.L. Laker and Henry J. Charlesworth }

\address{Department of Physics and Astronomy, University College London, Gower
Street, London WC1E 6BT, U.K.}
\begin{abstract}
We compute statistical properties of the stochastic entropy production
associated with the nonstationary transport of heat through a system
coupled to a time dependent nonisothermal heat bath. We study the
one-dimensional stochastic evolution of a bound particle in such an
environment by solving the appropriate Langevin equation numerically,
and by using an approximate analytic solution to the Kramers equation
to determine the behaviour of an ensemble of systems. We express the
total stochastic entropy production in terms of a relaxational or
nonadiabatic part together with two components of housekeeping entropy
production and determine the distributions for each, demonstrating
the importance of all three contributions for this system. We compare
the results with an approximate analytic model of the mean behaviour
and we further demonstrate that the total entropy production and the
relaxational component approximately satisfy detailed fluctuation
relations for certain time intervals. Finally, we comment on the resemblance
between the procedure for solving the Kramers equation and a constrained
extremisation, with respect to the probability density function, of
the spatial density of the mean rate of production of stochastic entropy.
\end{abstract}
\maketitle

\section{Introduction}

It is quite apparent that the macroscopic world largely operates in
an irreversible fashion, to the extent that the underlying time reversal
symmetry of the laws of physics is often obscured. Processes at the
macroscale typically evolve spontaneously in a specific direction
and not in reverse, unless driven to do so by external control. It
is very straightforward to list examples of irreversibility: heat
flow, chemical reaction, particle diffusion, decay of coherent motion,
inelastic collisions, brittle fracture and a host of other phenomena
of a dissipative character. The triumph of thermodynamics, from its
emergence in the 19th century up to the present day, is to interpret
all these phenomena as aspects of the second law.

Recent studies of irreversible processes at the microscopic level
have revealed a richer meaning of the traditional second law and of
the associated entropy production that quantifies the irreversibility
of a given process \cite{Evans94,Gallavotti95,jareqoriginal,Crooks99,Carberry04,Harris07,Ford-book2013}.
At microscopic scales, it is clear that a process can evolve both
forwards and backwards as a result of spontaneous fluctuations in
the system or its environment. A sequence of improbable but not impossible
collisions between molecules can drive a reaction from products back
into reactants for a short time, or a particle up instead of down
a concentration gradient. Developments in the thermodynamics of small
systems in recent years have made it possible to incorporate this
transitory behaviour into the same framework that accounts for the
much more clearly irreversible processes operating at the macroscale.
This broader viewpoint can be expressed through a framework of deterministic
dynamics \cite{Evans94,Gallavotti95}, or alternatively by using the
concepts of stochastic thermodynamics \cite{GCforstochastic,seifertoriginal,Harris07,seifertprinciples,sekimoto2},
where phenomenological noise is introduced into the dynamics of a
system coupled to an environment in order to account for and to quantify
dissipative behaviour.

The \emph{stochastic} entropy production that features in this latter
approach is defined in terms of the relative likelihood that a system
should evolve along a particular path or along its reverse. This quantity
has received considerable attention, particularly studies of the way
its statistics are governed by identities known as fluctuation relations
\cite{Harris07,SpinneyFordChapter13}. For example, while the stochastic
entropy production can be both positive and negative as a system evolves,
it satisfies an integral fluctuation relation which implies that its
expected rate of change is non-negative when averaged over many repeated
trials of the stochastic dynamics, real or imagined. When fluctuations
are small, departures from the second law are rare, but excursions
away from mean behaviour can be substantial for small systems or for
short processes, and the rules that govern this behaviour extend the
meaning of the second law at such scales.

Traditionally there has been just one measure of irreversibility:
thermodynamic entropy production, and many studies have investigated
its evolution in systems subject to dissipation, e.g. \cite{Bizarro08,Bizarro10}.
In stochastic thermodynamics, however, it has been possible to define
components of entropy production associated with different aspects
of irreversibility, each possessing specific properties \cite{adiabaticnonadiabatic0,adiabaticnonadiabatic1,adiabaticnonadiabatic2,SpinneyFord12a,SpinneyFord12b,SpinneyFord12c}.
For example, the irreversibility of the cooling of a saucepan of hot
soup differs somewhat from the irreversibility of the steady transport
of heat from a hot plate through the saucepan and into the surrounding
air that can maintain the soup at a desired temperature. Both are
associated with the production of thermodynamic entropy: in the first
case it may be described as relaxational, while in the second it has
been referred to as housekeeping production required to maintain a
steady state \cite{oono}. In stochastic thermodynamics each of these
modes of entropy production has been quantified in terms of the underlying
dynamics of a system coupled to an environment. Analysis has shown,
however, that two components are not always sufficient, and the housekeeping
element can separate into two parts, one of which has a transient
nature \cite{SpinneyFord12a,SpinneyFord12b,SpinneyFord12c}. Systems
driven by a time dependent environment under constraints that break
the principle of detailed balance in the underlying dynamics will
evolve irreversibly in a fashion characterised by three components
of stochastic entropy production. A similar conclusion was later reached
by \cite{Lee13} but with an alternative choice of representation.
In this study we investigate the statistics of the three contributions
for a simple case of thermal transport.

In Section \ref{sec:Stochastic-Thermodynamics-in} we introduce the
system of interest, a single particle performing underdamped Brownian
motion in a confining potential while coupled to a nonisothermal environment
characterised by a time and space dependent temperature. After defining
the components of stochastic entropy production, further details of
which are given in Appendix \ref{sec:Components-of-stochastic-1},
we employ in Section \ref{sec:Mean-stochastic-entropy} an approximate
solution to the Kramers equation describing the evolution of an ensemble
of such systems, derived in Appendix \ref{sec:Variational-solution-to},
to quantify the mean behaviour of each contribution. We note in Appendix
\ref{sec:Entropy-production} that the solution method resembles the
constrained extremisation of the spatial density of the mean rate
of production of stochastic entropy \cite{Kohler48,Ziman56,Cercignani-book2000},
analogous to Onsager's principle \cite{Onsager31}, although the interpretation
is not unproblematic. In Section \ref{sec:Distributions-of-stochastic}
we generate individual realisations of the motion using a Langevin
equation in order to obtain the distribution of fluctuations of stochastic
entropy production about the mean. Our particular aim is to demonstrate
the importance of the transient component of housekeeping entropy
production. We also demonstrate that as long as the friction coefficient
is not too small, the total entropy production as well as its relaxational
component satisfy detailed fluctuation relations for certain time
intervals and driving protocols, and in Appendix \ref{sec:Detailed-fluctuation}
we perform an analysis to provide an understanding of this behaviour.
This is a demonstration that an underlying exponential asymmetry in
the production and consumption of entropy can be made apparent as
long as the conditions are chosen carefully. In Section \ref{sec:Conclusions}
we present our conclusions.

\section{Stochastic thermodynamics in a nonisothermal environment\label{sec:Stochastic-Thermodynamics-in}}

A key aspect of stochastic thermodynamics is that it provides a link
between thermodynamic concepts, such as entropy production, and a
description of the mechanical evolution of a system. The stochastic
nature of the dynamics is important in that it ties in with an interpretation
of entropy production as the progressive loss of certainty in the
microscopic state of a system as time progresses. Such loss is perhaps
more fundamentally a consequence of a sensitivity to initial conditions
within a setting of deterministic dynamics, together with the difficulty
in preparing a system in a precise initial state, but introducing
phenomenological noise into a streamlined version of the dynamics
has a similar effect. Thermodynamics is the study of the behaviour
of a system in an environment where some of the features are specified
only approximately (this is particularly the case for the environment).
We therefore expect any modelling approach to have limited predictive
power. In stochastic thermodynamics it turns out that stochastic entropy
production, operationally defined in terms of certain energy exchanges,
embodies this predictive failure. It is striking to conclude that
microscopic uncertainty may essentially be measured using a thermometer,
and that an apparent determinism in the form of the second law for
large systems can emerge from a fundamentally underspecified dynamics.
It seems that one of the few matters about which we can be certain,
in such a situation, is that microscopic uncertainty should increase.

We focus our discussion on the one-dimensional (1-d) motion of a particle
coupled to a nonisothermal environment, described by the following
stochastic differential equations (SDEs):
\begin{eqnarray}
dx & = & vdt,\label{eq:1-1}\\
dv & = & -\gamma vdt+\frac{F(x)}{m}dt+\left(\frac{2kT_{r}(x,t)\gamma}{m}\right)^{1/2}dW,\label{eq:1a}
\end{eqnarray}
where $x$ and $v$ are the particle position and velocity, respectively,
$t$ is time, $\gamma$ is the friction coefficient, $F(x)$ is a
spatially dependent force field acting on the particle, assumed to
be related to a potential $\phi(x)$; $m$ is the particle mass, $T_{r}(x,t)$
is a space and time dependent environmental temperature and $dW$
is an increment in a Wiener process. Eq. (\ref{eq:1a}) is to be interpreted
using It$\bar{{\rm o}}$ rules of stochastic calculus \cite{Gardiner09,Celani12}.

We should note that heat baths are normally regarded as having static
thermal properties, so the time dependence of $T_{r}$ is to be interpreted
as the sequential decoupling and recoupling of the system to reservoirs
at slightly different temperatures. The effect of an evolving thermal
environment on a system can then be taken into account, retaining
the essential requirement that heat exchanges with the system should
not affect the properties of the environment. It should be noted that
a similar framework for discussing an evolving environmental temperature
in stochastic thermodynamics has recently been presented \cite{Brandner15}.
Indeed the above SDEs, often with a constant $T_{r}$ and in the overdamped
limit, have been used a starting point for discussing a great number
of characteristics of irreversible behaviour.

Following Seifert \cite{seifertoriginal}, stochastic entropy production
is defined as a measure of the probabilistic mechanical irreversibility
of the motion. The dynamics generate a trajectory $\vec{\boldsymbol{x}},\vec{\boldsymbol{v}}$
($\vec{\boldsymbol{x}}$ represents a function $x(t)$ in the time
interval $0\le t\le\tau$ and $\vec{\boldsymbol{v}}$ its time derivative)
under a `forward' driving protocol of force field and temperature
evolution. In Eq. (\ref{eq:1a}) the protocol is a specification of
the time dependence of $T_{r}$. The likelihood of the trajectory
is specified by a probability density function ${\rm P^{{\rm F}}}[\vec{\boldsymbol{x}},\vec{\boldsymbol{v}}]$
written as a product of the probability density of an initial microstate
$p(x=x(0),v=v(0),t=0)\equiv p_{{\rm start}}^{{\rm F}}[x(0),v(0)]$,
and a conditional probability density ${\cal P}$ for the subsequent
trajectory. The dynamics can also generate an \emph{antitrajectory}
initiated after an inversion of the particle velocity at time $\tau$,
and driven by a reversed time evolution of the force field and reservoir
temperature \cite{Ford-book2013,SpinneyFordChapter13,Ford15c}, until
a total time $2\tau$ has elapsed. Evolution in this interval $\tau\le t\le2\tau$
is described by a probability density ${\rm P}^{{\rm R}}[\vec{\boldsymbol{x}}^{\dagger},\vec{\boldsymbol{v}}^{\dagger}]$
for an antitrajectory $\vec{\boldsymbol{x}}^{\dagger},\vec{\boldsymbol{v}}^{\dagger}$
starting at $x(\tau),-v(\tau)$ and ending at $x(0),-v(0)$, with
the superscript R indicating that the potential and reservoir temperature
evolve backwards with respect to their evolution in the time interval
$0\le t\le\tau$ \cite{SpinneyFord12a,SpinneyFord12b,SpinneyFord12c}.
The total entropy production associated with the trajectory $\vec{\boldsymbol{x}},\vec{\boldsymbol{v}}$
is then defined by
\begin{equation}
\Delta s_{{\rm tot}}[\vec{\boldsymbol{x}},\vec{\boldsymbol{v}}]=\ln\left[\frac{{\rm P^{{\rm F}}}[\vec{\boldsymbol{x}},\vec{\boldsymbol{v}}]}{{\rm P}^{{\rm R}}[\vec{\boldsymbol{x}}^{\dagger},\vec{\boldsymbol{v}}^{\dagger}]}\right],\label{eq:2-1}
\end{equation}
and the key idea of stochastic thermodynamics is that after multiplication
by Boltzmann's constant and a procedure of averaging over all realisations
of the motion, this should correspond to the change in traditional
thermodynamic entropy associated with the forward process.

For the system under consideration, the stochastic entropy production
as the particle follows a trajectory evolves according to the SDE
\begin{equation}
d\Delta s_{{\rm tot}}=-d[\ln p(x,v,t)]-\frac{1}{kT_{r}(x,t)}d\left[\frac{mv^{2}}{2}\right]+\frac{F(x)dx}{kT_{r}(x,t)}.\label{eq:3-1}
\end{equation}
The derivation of this expression starting from the stochastic dynamics
in Eqs. (\ref{eq:1-1}) and (\ref{eq:1a}) is discussed in more detail
in Appendix \ref{sec:Components-of-stochastic-1}. The second and
third terms are negative increments in the kinetic and potential energy
of the particle over the time interval $dt$, both divided by the
local reservoir temperature. Together, they represent an increment
in the energy of the environment (a heat transfer $dQ_{r}$) divided
by the local temperature, therefore taking the form of an incremental
Clausius entropy production $d\Delta s_{r}=dQ_{r}/kT_{r}(x,t)$. The
first term in Eq. (\ref{eq:3-1}) is the stochastic entropy production
associated with the particle over the time interval. Seifert defined
a stochastic system entropy $s_{{\rm sys}}=-\ln p(x,v,t)$ in terms
of the phase space probability density function $p$ generated by
the stochastic dynamics \cite{seifertoriginal}, such that we can
write $d\Delta s_{{\rm tot}}=d\Delta s_{{\rm sys}}+d\Delta s_{r}.$
As the particle follows a trajectory, it moves through a probability
density function $p(x,v,t)$ that represents all the possible paths
that could have been followed, and the system entropy production emerges
from a comparison between the actual event and this range of possible
behaviour. The evaluation of $\Delta s_{{\rm tot}}$ for a specific
realisation of the motion therefore requires us to determine the probability
density function (pdf) by solving the appropriate Kramers equation
\cite{Risken89}
\begin{equation}
\frac{\partial p}{\partial t}={\cal L}p=-\frac{\partial J_{v}^{{\rm ir}}}{\partial v}-v\frac{\partial p}{\partial x}-\frac{F}{m}\frac{\partial p}{\partial v},\label{eq:fpe}
\end{equation}
corresponding to the SDEs in Eqs. (\ref{eq:1-1}) and (\ref{eq:1a}),
where $J_{v}^{{\rm ir}}=-\gamma vp-\partial(D_{v}p)/\partial v$,
with $D_{v}=\gamma kT_{r}(x,t)/m$, is the irreversible probability
current for this system, responsible for the growth of uncertainty
and hence mean stochastic entropy production.

In spite of the fluctuating nature of the total stochastic entropy
production, the expectation of this quantity is non-negative. This
may be expressed as $d\langle\Delta s_{{\rm tot}}\rangle=d\langle\Delta s_{{\rm sys}}\rangle+d\langle\Delta s_{r}\rangle\ge0$
where the brackets denote an average over the distributions of system
coordinates at the beginning and end of the incremental time period.
Note that for economy the qualifier `stochastic' is henceforth to
be implied rather than stated when referring to entropy production.

We now separate the entropy production into components, each with
a particular character, along the lines of initial developments by
Van den Broeck and Esposito \cite{adiabaticnonadiabatic0,adiabaticnonadiabatic1,adiabaticnonadiabatic2}
and extended by Spinney and Ford \cite{SpinneyFord12a,SpinneyFord12b,SpinneyFord12c},
using a framework suggested by Oono and Paniconi \cite{oono}. The
total entropy production may be written as three terms \cite{SpinneyFord12a,SpinneyFord12b}
\begin{equation}
d\Delta s_{{\rm tot}}=d\Delta s_{1}+d\Delta s_{2}+d\Delta s_{3},\label{eq:5a}
\end{equation}
with the $\Delta s_{1}$ and $\Delta s_{2}$ components defined in
terms of ratios of probabilities that specific trajectories are taken
by the system, in a manner similar to Eq. (\ref{eq:2-1}). Details
are to be found elsewhere \cite{SpinneyFord12a,SpinneyFord12b,SpinneyFord12c}
and in Appendices \ref{sec:Components-of-stochastic-1} and \ref{sec:Detailed-fluctuation}.
Note that there is no implication of a one-to-one correspondence between
the $\Delta s_{1-3}$ and the three terms in Eq. (\ref{eq:3-1}).

The evolution of the average values of the components may be related
to the transient and stationary system pdfs ($p$ and $p_{{\rm st}}$,
respectively) according to
\begin{eqnarray}
\frac{d\langle\Delta s_{1}\rangle}{dt} & = & \int dxdv\;\frac{p}{D_{v}}\left(\frac{J_{v}^{{\rm ir}}}{p}-\frac{J_{v}^{{\rm ir,st}}}{p_{{\rm st}}}\right)^{2}\ge0,\label{S1av2}\\
\frac{d\langle\Delta s_{2}\rangle}{dt} & = & \int dxdv\;\frac{p}{D_{v}}\left(\frac{J_{v}^{{\rm ir,st}}(x,-v)}{p_{{\rm st}}(x,-v)}\right)^{2}\ge0,\label{eq:s2av}\\
\frac{d\langle\Delta s_{3}\rangle}{dt} & = & -\int dxdv\;\frac{\partial p}{\partial t}\ln\left[\frac{p_{{\rm st}}(x,v)}{p_{{\rm st}}(x,-v)}\right],\label{eq:s3av}
\end{eqnarray}
where ${\cal L}p_{{\rm st}}=0$, and $J_{v}^{{\rm ir,st}}=-\gamma vp_{{\rm st}}-\partial(D_{v}p_{{\rm st}})/\partial v$
is the irreversible probability current in the stationary state. The
mean rate of total entropy production is
\begin{equation}
\frac{d\langle\Delta s_{{\rm tot}}\rangle}{dt}=\int dxdv\;\frac{(J_{v}^{{\rm ir}})^{2}}{D_{v}p}\ge0.\label{eq:stot}
\end{equation}
  The three contributions to the total entropy production can be
interpreted as follows. $\Delta s_{1}$ is the principal relaxational
entropy production associated with the approach of a system towards
a stationary state. Its average over all possible realisations of
the motion, namely $\langle\Delta s_{1}\rangle$, increases monotonically
with time until stationarity is reached, since $d\langle\Delta s_{1}\rangle/dt\to0$
as $p\to p_{{\rm st}}$. Esposito and Van den Broeck \cite{adiabaticnonadiabatic0,adiabaticnonadiabatic1,adiabaticnonadiabatic2}
denoted it the nonadiabatic entropy production.

$\Delta s_{3}$ is also associated with relaxation, but in contrast
to $\Delta s_{1}$ no definite sign can be attached to $d\langle\Delta s_{3}\rangle/dt$.
However, if the stationary pdf is velocity symmetric $\Delta s_{3}$
is identically zero. Since a velocity asymmetric stationary pdf is
typically associated with breakage of a principle of detailed balance
in the stochastic dynamics \cite{adiabaticnonadiabatic0,adiabaticnonadiabatic1,adiabaticnonadiabatic2},
this component arises in situations where there is a nonequilibrium
stationary state involving velocity variables. It was designated the
transient housekeeping entropy production by Spinney and Ford \cite{SpinneyFord12a}.

$\Delta s_{2}$ is also associated with a nonequilibrium stationary
state, since its average rate of change in Eq. (\ref{eq:s2av}) requires
a non-zero current $J_{v}^{{\rm ir,st}}$ in the stationary state.
The mean entropy production rate in the stationary state is represented
by $d\langle\Delta s_{2}\rangle/dt$ alone, and this is non-zero only
if the stationary current $J_{v}^{{\rm ir,st}}$ is non-zero. Esposito
and Van den Broeck referred to $\Delta s_{2}$ as the adiabatic entropy
production and considered it in the context of the dynamics of spatial
coordinates, and Spinney and Ford, who considered velocity variables
as well, denoted it the generalised housekeeping entropy production.

For a nonisothermal, time dependent environment we expect all three
kinds of entropy production to take place. The mean rates of production
for each component are examined next, and in Section \ref{sec:Distributions-of-stochastic}
we shall consider the distributions of fluctuations about the mean.

\section{Mean stochastic entropy production\label{sec:Mean-stochastic-entropy}}

We are concerned with the 1-d Brownian motion of a particle in a potential
under the influence of a background temperature $T_{r}$ that varies
in space and time. The pdf $p(x,v,t)$ evolves according to Eq. (\ref{eq:fpe})
subject to a requirement that $p$, $\partial p/\partial v$ and $\partial p/\partial x$
all vanish as $x,v\to\pm\infty$ for all $t$. We shall use an established
perturbative method \cite{Hirschfelder54,Lebowitz60,Cercignani-book2000}
to obtain an approximate expression for $p$ to leading order in the
inverse friction coefficient.

Integration of the Kramers equation with respect to $v$ yields the
continuity equation
\begin{equation}
\frac{\partial\rho}{\partial t}+\frac{\partial(\rho\bar{v})}{\partial x}=0,\label{eq:3}
\end{equation}
where we define $\rho(x,t)=\int dvp$ and $\rho\overline{v^{n}}(x,t)=\int dvv^{n}p$,
and multiplication by $v$ followed by integration gives
\begin{equation}
\frac{\partial(\rho\bar{v})}{\partial t}+\frac{\partial(\rho\overline{v^{2}})}{\partial x}-\frac{F}{m}\rho=-\gamma\rho\bar{v},\label{eq:4}
\end{equation}
which is a momentum transport equation.  We represent the pdf in
the form $p=f(1+\psi)$ with
\begin{equation}
f(x,v,t)=\rho\left(\frac{m}{2\pi kT_{r}}\right)^{\frac{1}{2}}\exp\left(-\frac{m\left(v-\bar{v}\right)^{2}}{2kT_{r}}\right),\label{eq:6}
\end{equation}
such that $\int dvf\psi=0$ and $\int dvvf\psi=0$. We further simplify
the situation by requiring that the time dependence in the pdf is
confined to the distribution over velocity. The spatial pdf $\rho$
is therefore time independent which in turn implies that the mean
velocity $\bar{v}$ is zero, according to Eq. (\ref{eq:3}). We study
situations where the background temperature is driven in a cyclic
manner with $T_{r}(x,t)=T_{r}^{0}(x)[1+g(t)]$, but not so violently
that $\rho$ is significantly disturbed from its profile when $g=0$.
 We write $\rho\overline{v^{2}}(x,t)=\rho\bar{v}^{2}+\rho kT_{r}/m+\int dvv^{2}f\psi$,
and  anticipating that the final term is of order $\gamma^{-1}$
we deduce that $\rho\overline{v^{2}}\approx\rho kT_{r}^{0}/m$ for
the stationary case where $\bar{v}=0$ and $g=0$, so that Eq. (\ref{eq:4})
reduces to
\begin{equation}
k\frac{\partial(\rho_{{\rm st}}T_{r}^{0})}{\partial x}-F\rho_{{\rm st}}\approx0,\label{eq:6-1}
\end{equation}
in which case
\begin{equation}
\rho_{{\rm st}}(x)\propto\frac{1}{T_{r}^{0}(x)}\exp\left(\int_{0}^{x}dx^{\prime}\frac{F(x^{\prime})}{kT_{r}^{0}(x^{\prime})}\right),\label{eq:6a}
\end{equation}
is the approximation we shall employ for the spatial distribution
$\rho$.

The Kramers equation is
\begin{equation}
\frac{\partial p}{\partial t}+v\frac{\partial p}{\partial x}+\left(\frac{F}{m}-\gamma\bar{v}\right)\frac{\partial p}{\partial v}=\frac{kT_{r}\gamma}{m}\frac{\partial}{\partial v}\left(f\frac{\partial\psi}{\partial v}\right),\label{eq:7}
\end{equation}
and we set $\bar{v}=0$ and expand the distribution $p=f(1+\psi)$
as a series in $\gamma^{-1}$. The leading term $p\approx f$ is independent
of $\gamma$ and we write
\begin{equation}
\psi=\gamma^{-1}\psi_{1}+\gamma^{-2}\psi_{2}+\cdots,\label{eq:7a}
\end{equation}
with each contribution $\psi_{i}$ satisfying $\int dvf\psi_{i}=0$
and $\int dvvf\psi_{i}=0$. Gathering all terms in Eq. (\ref{eq:7})
of order zero in $\gamma^{-1}$ and setting $\bar{v}=0$ leads to
\begin{equation}
\frac{\partial f}{\partial t}+v\frac{\partial f}{\partial x}+\frac{F}{m}\frac{\partial f}{\partial v}=\frac{kT_{r}\gamma}{m}\frac{\partial}{\partial v}\left(f\frac{\partial(\gamma^{-1}\psi_{1})}{\partial v}\right),\label{eq:8}
\end{equation}
and by solving this for $\psi_{1}$, the representation of $p$ that
emerges will be correct to first order in $\gamma^{-1}$.

It is possible to obtain a solution to Eq. (\ref{eq:8}) using a variational
procedure that is described in more detail in Appendix \ref{sec:Variational-solution-to},
and to argue that the identification of $p$ resembles a principle
of constrained extremisation of the spatial density of the mean rate
of entropy production specified to first order in the inverse friction
coefficient (see Appendix \ref{sec:Entropy-production}). Here it
is sufficient to state that Eq. (\ref{eq:8}) is satisfied by Eq.
(\ref{eq:6}) with $\bar{v}=0$ and with

\begin{equation}
\gamma^{-1}\psi_{1}=\psi_{0}+av+bv^{2}+cv^{3},\label{eq:14a-1}
\end{equation}
 where
\begin{eqnarray}
a & = & \frac{1}{2\gamma T_{r}}\frac{\partial T_{r}}{\partial x},\qquad b=-\frac{m}{4\gamma kT_{r}^{2}}\frac{\partial T_{r}}{\partial t},\nonumber \\
c & = & -\frac{m}{6\gamma kT_{r}^{2}}\frac{\partial T_{r}}{\partial x},\qquad\psi_{0}=\frac{1}{4\gamma T_{r}}\frac{\partial T_{r}}{\partial t},\label{eq:16-1}
\end{eqnarray}
such that upon insertion of Eq. (\ref{eq:14a-1}) into Eq. (\ref{eq:8})
the coefficients of terms proportional to powers of $v$ vanish. The
evolving pdf is therefore specified by
\begin{eqnarray}
 &  & p(x,v,t)\;\approx\Biggl[1+\frac{1}{4\gamma T_{r}}\frac{\partial T_{r}}{\partial t}+\frac{1}{2\gamma T_{r}}\frac{\partial T_{r}}{\partial x}v\nonumber \\
 &  & \quad\qquad-\frac{m}{4\gamma kT_{r}^{2}}\frac{\partial T_{r}}{\partial t}v^{2}-\frac{m}{6\gamma kT_{r}^{2}}\frac{\partial T_{r}}{\partial x}v^{3}\Biggr]\nonumber \\
 &  & \quad\qquad\times\rho_{{\rm st}}(x)\left(\frac{m}{2\pi kT_{r}}\right)^{\frac{1}{2}}\exp\left(-\frac{mv^{2}}{2kT_{r}}\right),\label{eq:38-1}
\end{eqnarray}
to first order in $\gamma^{-1}$, and the stationary pdf for a given
temperature profile $T_{r}$ is
\begin{eqnarray}
p_{{\rm st}}(x,v)\; & \approx & \Biggl[1+\frac{1}{2\gamma T_{r}}\frac{\partial T_{r}}{\partial x}v-\frac{m}{6\gamma kT_{r}^{2}}\frac{\partial T_{r}}{\partial x}v^{3}\Biggr]\nonumber \\
 & \times & \rho_{{\rm st}}(x)\left(\frac{m}{2\pi kT_{r}}\right)^{\frac{1}{2}}\exp\left(-\frac{mv^{2}}{2kT_{r}}\right).\label{eq:38-2}
\end{eqnarray}

Now we can evaluate the mean rate of change of each component of entropy
production. Most straightforwardly, from Eqs. (\ref{eq:s3av}) and
(\ref{eq:38-2}) we have
\begin{equation}
\frac{d\langle\Delta s_{3}\rangle}{dt}\!=\!-\int\! dxdv\,\frac{\partial p}{\partial t}\ln\left[\frac{1+\frac{1}{2\gamma T_{r}}\frac{\partial T_{r}}{\partial x}v-\frac{m}{6\gamma kT_{r}^{2}}\frac{\partial T_{r}}{\partial x}v^{3}}{1-\frac{1}{2\gamma T_{r}}\frac{\partial T_{r}}{\partial x}v+\frac{m}{6\gamma kT_{r}^{2}}\frac{\partial T_{r}}{\partial x}v^{3}}\right].\label{eq:61}
\end{equation}
The logarithm is an odd function of $v$ and its leading term is proportional
to $\gamma^{-1}$, and furthermore, according to Eq. (\ref{eq:38-1}),
$\partial p/\partial t$ is an even function of $v$ to zeroth order
in $\gamma^{-1}$, so we conclude that $\langle\Delta s_{3}\rangle\approx0$
to order $\gamma^{-1}$.

The irreversible probability current is
\begin{eqnarray}
J_{v}^{{\rm ir}} & = & \frac{k}{2m}\Biggl[\frac{\partial T_{r}}{\partial x}-\frac{m}{kT_{r}}\frac{\partial T_{r}}{\partial t}v-\frac{m}{kT_{r}}\frac{\partial T_{r}}{\partial x}v^{2}\Biggr]\nonumber \\
 & \times & \rho_{{\rm st}}(x)\left(\frac{m}{2\pi kT_{r}}\right)^{\frac{1}{2}}\exp\left(-\frac{mv^{2}}{2kT_{r}}\right),\label{eq:62}
\end{eqnarray}
 so that from Eqs. (\ref{eq:s2av}) and (\ref{eq:62}) we have
\begin{eqnarray}
\frac{d\langle\Delta s_{2}\rangle}{dt} & \approx & \int dxdv\;\frac{kf}{4\gamma mT_{r}}\left(\frac{\partial T_{r}}{\partial x}\Biggl[1-\frac{m}{kT_{r}}v^{2}\Biggr]\right)^{2}\nonumber \\
 & = & \int dx\;\frac{k\rho_{{\rm st}}}{2\gamma mT_{r}}\left(\frac{\partial T_{r}}{\partial x}\right)^{2},\label{eq:63}
\end{eqnarray}
to first order in $\gamma^{-1}$. Finally, from Eqs. (\ref{S1av2})
and (\ref{eq:62}) we deduce that
\begin{eqnarray}
\frac{d\langle\Delta s_{1}\rangle}{dt} & \approx & \int dxdv\;\frac{fm}{\gamma kT_{r}}\left[\frac{1}{2T_{r}}\frac{\partial T_{r}}{\partial t}v\right]^{2}\nonumber \\
 & = & \int dx\;\frac{\rho_{{\rm st}}}{4\gamma T_{r}^{2}}\left(\frac{\partial T_{r}}{\partial t}\right)^{2},\label{eq:64}
\end{eqnarray}
such that
\begin{equation}
\frac{d\langle\Delta s_{{\rm tot}}\rangle}{dt}\approx\int dx\:\frac{k}{2m\gamma T_{r}}\rho_{{\rm st}}\left[\frac{m}{2kT_{r}}\left(\frac{\partial T_{r}}{\partial t}\right)^{2}+\left(\frac{\partial T_{r}}{\partial x}\right)^{2}\right],\label{eq:65}
\end{equation}
which can also be obtained by the direct insertion of Eqs. (\ref{eq:38-1})
and (\ref{eq:62}) into Eq. (\ref{eq:stot}). The relaxational and
housekeeping (nonadiabatic and adiabatic in alternative terminology)
components of the mean rate of total entropy production, to leading
order in $\gamma^{-1}$, are clearly never negative and can be seen
to arise from the temporal and spatial dependence, respectively, of
the environmental temperature. For $\partial T_{r}/\partial t=0$,
Eq. (\ref{eq:65}) reduces to an expression employed previously in
studies of entropy production in a time independent nonisothermal
system \cite{Stolovitzky98,SpinneyFord12b}.

\section{Distributions of stochastic entropy production\label{sec:Distributions-of-stochastic}}

\begin{figure}
\begin{centering}
\includegraphics[width=1\columnwidth]{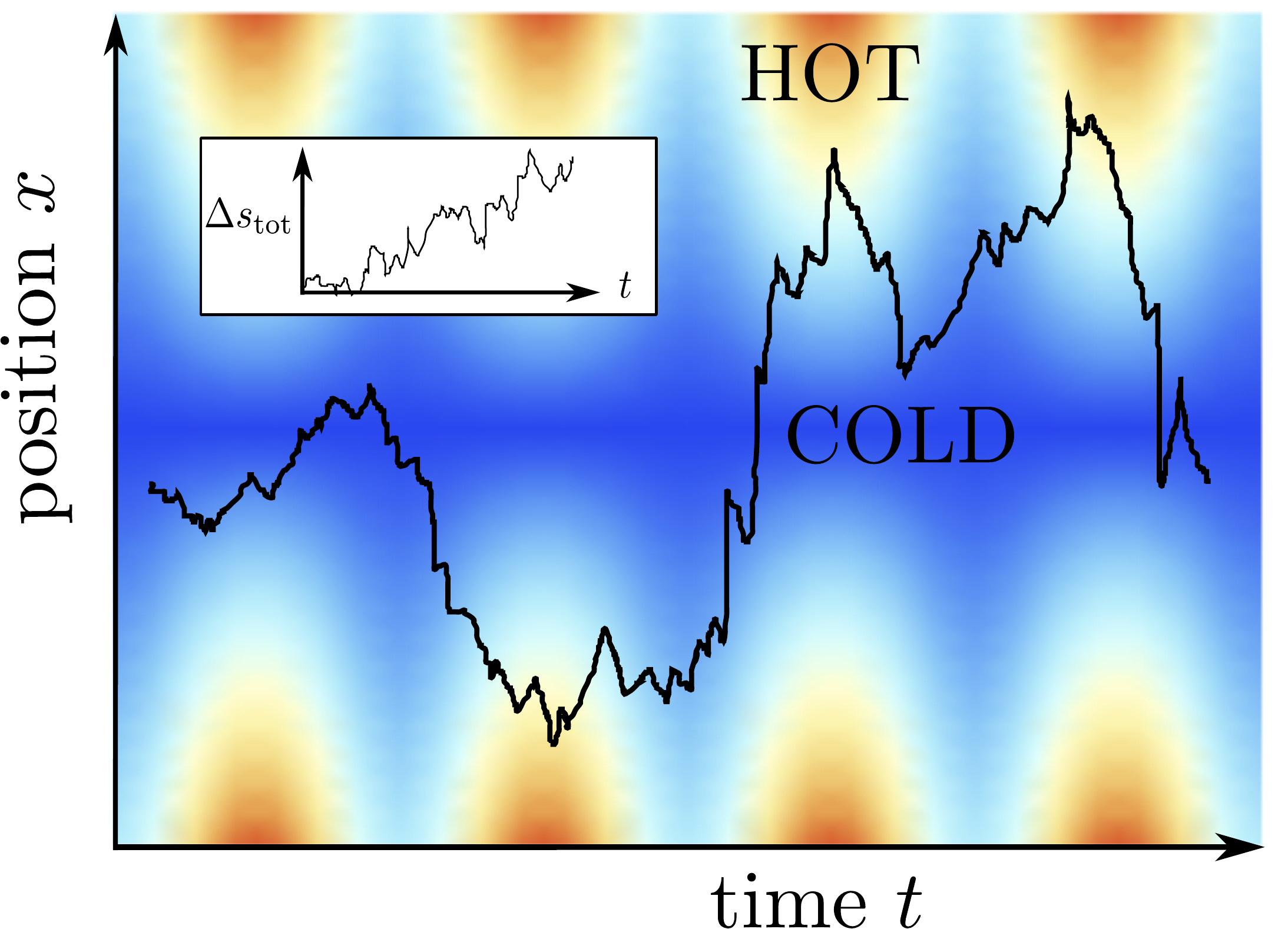}
\par\end{centering}

\protect\caption{Spatial trajectory of a particle according to stochastic dynamics
in the presence of an environment where the temperature varies in
time and space, illustrated by the text and background colours. The
evolution of the stochastic entropy production $\Delta s_{{\rm tot}}$
associated with the Brownian trajectory is sketched in the inset.
\label{fig:1}}
\end{figure}

We now turn our attention to fluctuations in the production of stochastic
entropy away from the mean behaviour determined in the last section.
We solve the SDEs (\ref{eq:1-1}) and (\ref{eq:1a}) to generate a
trajectory of particle position and velocity and then insert the pdf
specified in Eq. (\ref{eq:38-1}) into Eq. (\ref{eq:3-1}) to obtain
the associated evolution of total entropy production $\Delta s_{{\rm tot}}$.
This is illustrated in Figure \ref{fig:1} where a particle follows
a Brownian trajectory while coupled to an environment with a temperature
that varies in space and time, represented by the background colours.
The evolution of $\Delta s_{{\rm tot}}$ is stochastic, but with a
distinct upward trend.

\begin{figure}
\centering{} \includegraphics[width=1\columnwidth]{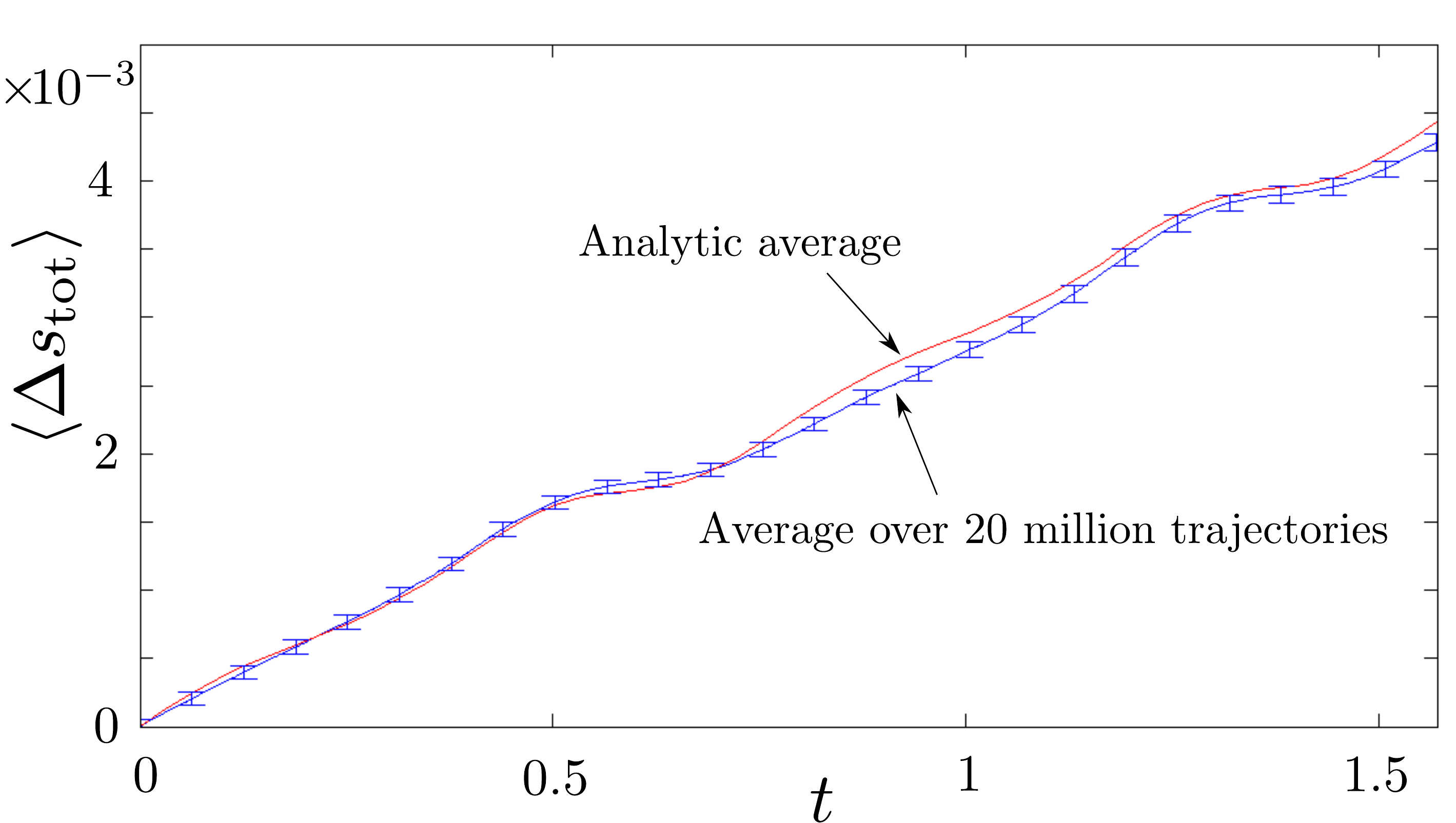} \protect\caption{A comparison between the mean total stochastic entropy production
obtained from the integral of the approximate Eq. (\ref{eq:65}) and
the total stochastic entropy production averaged over $2\times10^{7}$
numerically generated particle trajectories, each with 40000 timesteps,
shown with error bars. The time interval represents two cycles of
temperature variation, preceded in the simulations by one cycle to
establish an evolving pdf similar to Eq. (\ref{eq:38-1}).\label{fig:stot}}
\end{figure}

We choose simple forms of the potential and thermal background that
the particle experiences. We consider a harmonic force $F(x)=-\kappa x$
where $\kappa$ is a spring constant, and an environmental temperature
that varies in space and time according to
\begin{equation}
T_{r}(x,t)=T_{0}[1+\kappa_{T}(t)x^{2}/(2kT_{0})],\label{eq:39}
\end{equation}
where $T_{0}$ is a constant and the time dependence is specified
by $\kappa_{T}(t)=\kappa_{T}^{0}+B\sin\omega t>0$ with constant $B$.
Four cycles of such behaviour are sketched in Figure \ref{fig:1}.
The environment is hotter as the distance from the particle tether
point increases, and the spatial temperature profile varies sinusoidally
with time. We expect heat to be carried by the particle, on average,
from sources located away from the tether point towards sinks situated
near the centre of the motion. The average rate of flow of heat should
be affected by the time dependence of $T_{r}$.

This choice of profile implies that $T_{r}^{0}(x)=T_{0}[1+\kappa_{T}^{0}x^{2}/(2kT_{0})]$
and we evaluate the integral
\begin{eqnarray}
\int_{0}^{x}dx\frac{F(x)}{kT_{r}^{0}(x)} & = & -\int_{0}^{x}dx\frac{\kappa x}{kT_{0}[1+\kappa_{T}^{0}x^{2}/(2kT_{0})]}\nonumber \\
 & = & -\left(\kappa/\kappa_{T}^{0}\right)\ln[1+\kappa_{T}^{0}x^{2}/(2kT_{0})],\label{eq:40}
\end{eqnarray}
such that the normalised stationary spatial pdf according to Eq. (\ref{eq:6a})
is
\begin{equation}
\rho_{{\rm st}}(x)=\left(\frac{\kappa_{T}^{0}}{2\pi kT_{0}}\right)^{\frac{1}{2}}\frac{\Gamma\left(1+\kappa/\kappa_{T}^{0}\right)}{\Gamma\left(\frac{1}{2}+\kappa/\kappa_{T}^{0}\right)}\left[1+\frac{\kappa_{T}^{0}}{2kT_{0}}x^{2}\right]^{-1-\kappa/\kappa_{T}^{0}}\label{eq:42}
\end{equation}
and furthermore we can write
\begin{equation}
\frac{\partial T_{r}}{\partial x}=\kappa_{T}(t)\frac{x}{k},\label{eq:43}
\end{equation}
and
\begin{equation}
\frac{\partial T_{r}}{\partial t}=\frac{d\kappa_{T}(t)}{dt}\frac{x^{2}}{2k},\label{eq:44}
\end{equation}
which fully specifies the evolution of $\langle\Delta s_{{\rm tot}}\rangle$
given in Eq. (\ref{eq:65}). A similar system with time-independent
$\kappa_{T}$ was examined in \cite{FordEyre15} for the purpose of
deriving work relations under nonisothermal conditions.

For the numerical computation of the total entropy production we integrate
\begin{equation}
d\Delta s_{{\rm tot}}=-d[\ln p]-\frac{m}{2kT_{r}}d\left[v^{2}\right]-\frac{\kappa x}{kT_{r}}dx,\label{eq:33-1}
\end{equation}
along with the SDEs for $x$ and $v$, and the components of entropy
production evolve \cite{SpinneyFord12b} according to

\begin{eqnarray}
d\Delta s_{1} & = & -d[\ln p]+d[\ln p_{{\rm st}}],\label{eq:33-2}\\
d\Delta s_{3} & = & -d[\ln p_{{\rm st}}(x,v)]+d[\ln(p_{{\rm st}}(x,-v)],\label{eq:33-3}
\end{eqnarray}
together with $d\Delta s_{2}=d\Delta s_{{\rm tot}}-d\Delta s_{1}-d\Delta s_{3}$.

We select initial coordinates from the stationary pdf $p_{{\rm st}}(x,v)$
specified by $T_{r}^{0}(x)$, the temperature profile at $t=0$, and
evolve the system over a time interval $0\le t\le6\pi/\omega$ corresponding
to three cycles of variation in the temperature profile, with parameters
$\omega=8$, $m=1$, $kT_{0}=1$, $\kappa=1$, $\kappa_{T}^{0}=0.5$,
$B=0.2$, and $\gamma=60$. The short relaxation time $\gamma^{-1}$
relative to the cycle period $2\pi/\omega$ ensures that the dynamics
do not depart very far from the overdamped limit such that the expressions
for $p$ and $p_{{\rm st}}$ given in Eqs. (\ref{eq:38-1}) and (\ref{eq:38-2})
are reasonably accurate.

\begin{figure}
\centering{} \includegraphics[width=1\columnwidth]{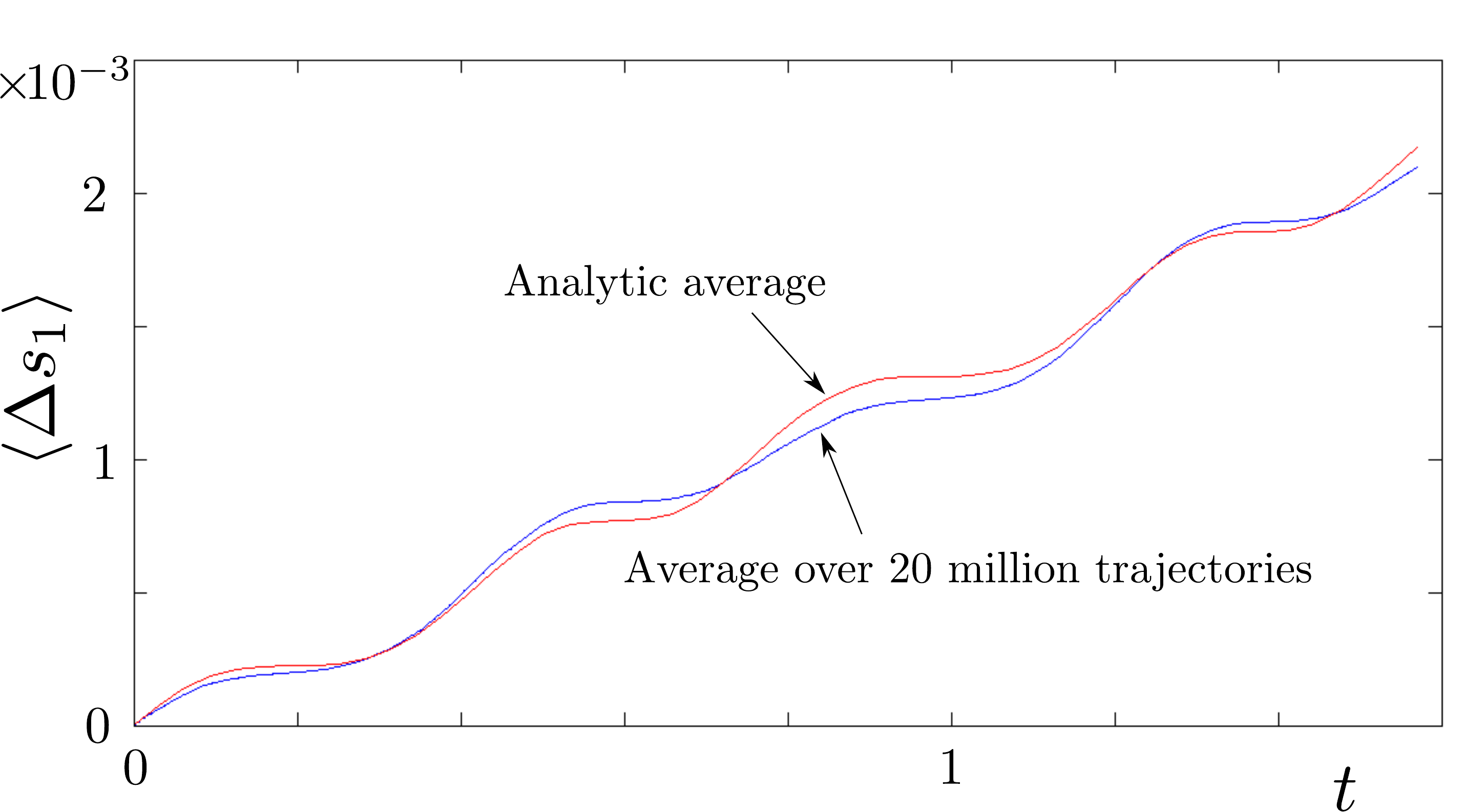} \protect\caption{Comparison of the average of $\Delta s_{1}$ predicted analytically
and calculated from the simulation of $2\times10^{7}$ particle trajectories
over a time interval of two temperature cycles preceded by one cycle
to establish the periodic stationary state. \label{fig:s1}}
\end{figure}

\begin{figure}
\centering{} \includegraphics[width=1\columnwidth]{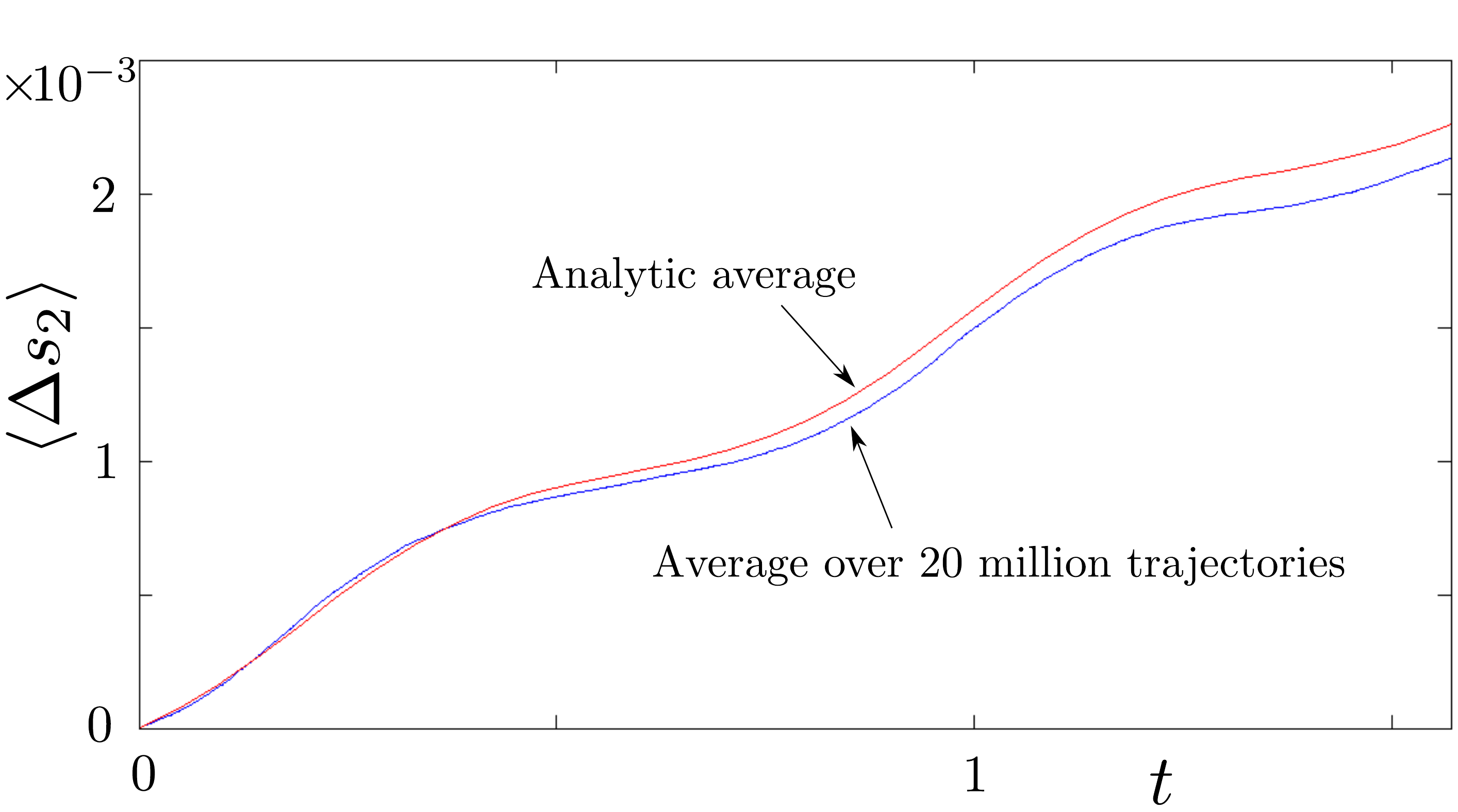} \protect\caption{As Figure \ref{fig:s1} but for the average of $\Delta s_{2}$.\label{fig:s2}}
\end{figure}

The interval is divided into 40000 timesteps of length $\delta t=5.89\times10^{-5}$
and samples of entropy production are generated from $2\times10^{7}$
realisations of the Brownian motion. The system relaxes quickly into
a periodic stationary state but we disregard behaviour taking place
in the first cycle $0\le t\le2\pi/\omega$ and focus our attention
on entropy production during the second and third cycles.

We gauge the quality of the numerical calculations by checking that
$\langle\Delta s_{{\rm tot}}\rangle$ matches the evolution obtained
from integrating the approximate analytical expression (\ref{eq:65}).
Statistical uncertainty in the numerical results is assessed by blocking
the realisations into 40 subsets, and the resulting error bars in
Figure \ref{fig:stot} show that the accuracy of the numerical approach
is satisfactory. Similar conclusions are reached by determining the
evolution of $\langle\Delta s_{1}\rangle$ by the analytic and numerical
routes, illustrated in Figure \ref{fig:s1} and a similar procedure
for $\langle\Delta s_{2}\rangle$ in Figure \ref{fig:s2}. The different
character of these two components of entropy production is apparent.
There are two bursts of relatively rapid mean production of $\Delta s_{1}$
per cycle. The system responds to the raising and lowering of the
temperature profile and relaxational entropy generation is associated
with both. In contrast, the time development of $\langle\Delta s_{2}\rangle$
more closely matches the periodicity of the temperature cycle, since
it is a reflection of the entropy production that would characterise
a stationary state for a given nonisothermal profile.

\begin{figure}
\centering{} \includegraphics[width=1\columnwidth]{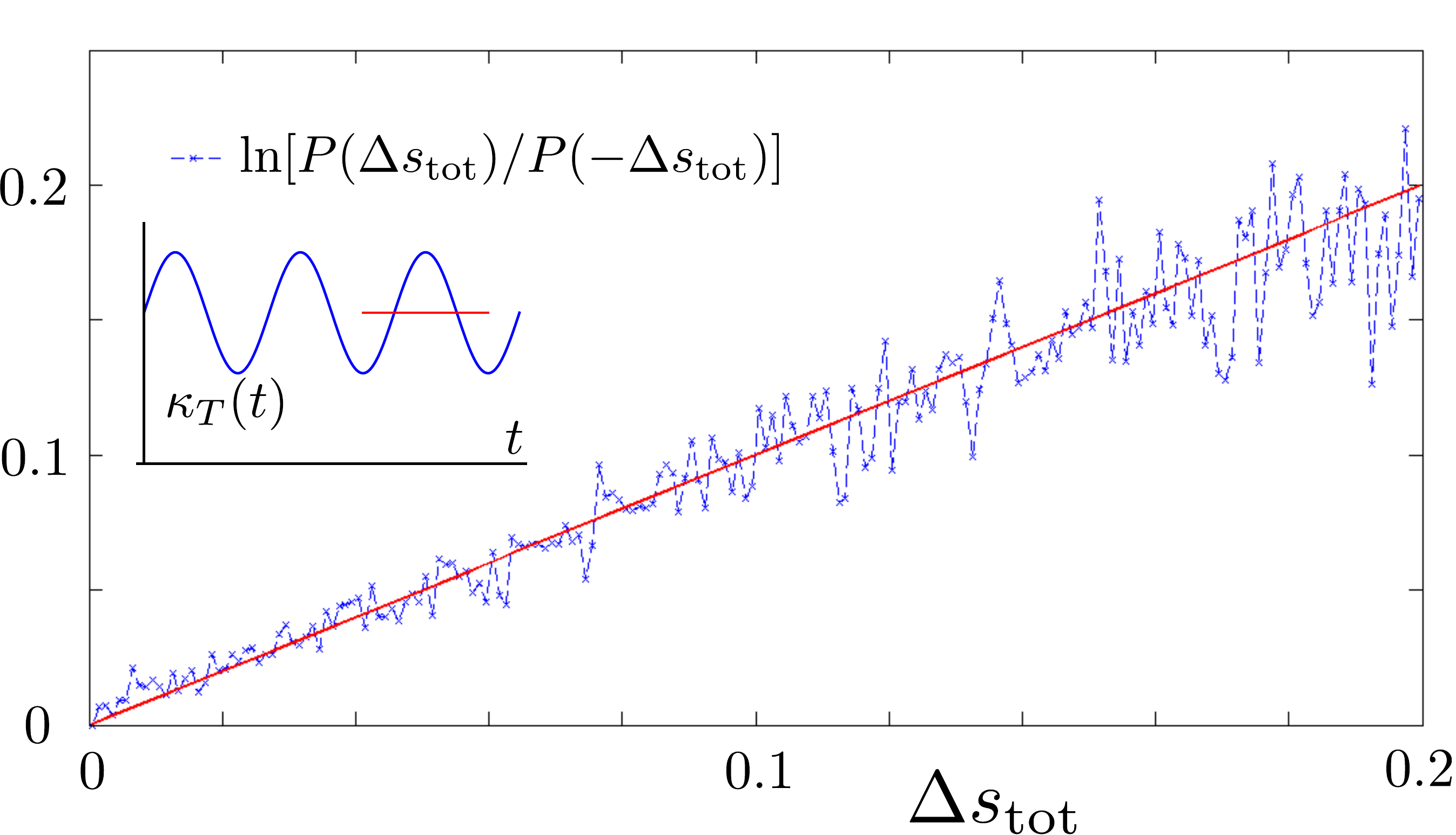} \protect\caption{Numerical verification of an approximate detailed fluctuation relation
for the pdf of total stochastic entropy production $\Delta s_{{\rm tot}}$
in the time interval $7\pi/2\omega\le t\le11\pi/2\omega$ (shown as
a bar in the inset), driven by the change in temperature profile according
to the evolving $\kappa_{T}(t)$. Values of $\Delta s_{{\rm tot}}$
between $\pm0.2$ from the simulation of $2\times10^{7}$ particle
trajectories are collected into 400 bins, and the straight line represents
the outcome expected in the absence of sampling errors.\label{fig:dftstot}}
\end{figure}

The $\Delta s_{1}$ and $\Delta s_{2}$ components of entropy production,
as well as $\Delta s_{{\rm tot}}$, satisfy an integral fluctuation
relation $\langle\exp(-\Delta s_{i})\rangle=1$ by construction for
any elapsed time interval \cite{adiabaticnonadiabatic0,SpinneyFord12a}.
We can demonstrate further that a detailed fluctuation relation $P(\Delta s_{{\rm tot}})=P(-\Delta s_{{\rm tot}})\exp(\Delta s_{{\rm tot}})$
appears to be satisfied by $\Delta s_{{\rm tot}}$ for certain time
intervals, as illustrated in Figure \ref{fig:dftstot}. We have chosen
conditions where the initial and final system pdfs are the same, which
will be the case for an interval that is a multiple of the cycle period
$2\pi/\omega$ once the system has adopted a periodic stationary state;
and for which the evolution of the environmental temperature profile
is symmetric about the midpoint of the time interval, for example
$7\pi/2\omega\le t\le11\pi/2\omega$ indicated by the horizontal bar
in the inset shown in Figure \ref{fig:dftstot}. These are circumstances
where the total entropy production in a system described by spatial
coordinates alone \cite{Harris07,SpinneyFordChapter13} is expected
to satisfy a detailed fluctuation relation. The backward version of
the process in this time interval is identical to the forward version.
Detailed fluctuation relations relate distributions of entropy production
in forward and backward processes but here the two are synonymous.
For time intervals where this is not the case, for example $3\pi/\omega\le t\le5\pi/\omega$,
the distribution of total entropy production will not satisfy a detailed
fluctuation relation.

For systems that possess velocity coordinates, a detailed fluctuation
relation for $\Delta s_{{\rm tot}}$ will be valid if, additionally,
the initial pdf for the backward process is the time-reversed version
of the initial pdf for the forward process. For such a relation to
hold for our system, the pdf at the beginning and end of the cycle
should be velocity symmetric, as demonstrated in Appendix \ref{sub:Detailed-fluctuation-relation 1}.
This condition is not in general satisfied for a nonequilibrium system
described by underdamped dynamics, but if the friction coefficient
is not too small, the velocity asymmetry in Eq. (\ref{eq:38-1}) is
slight and the detailed fluctuation relation should hold to a good
approximation.

\begin{figure}
\centering{} \includegraphics[width=1\columnwidth]{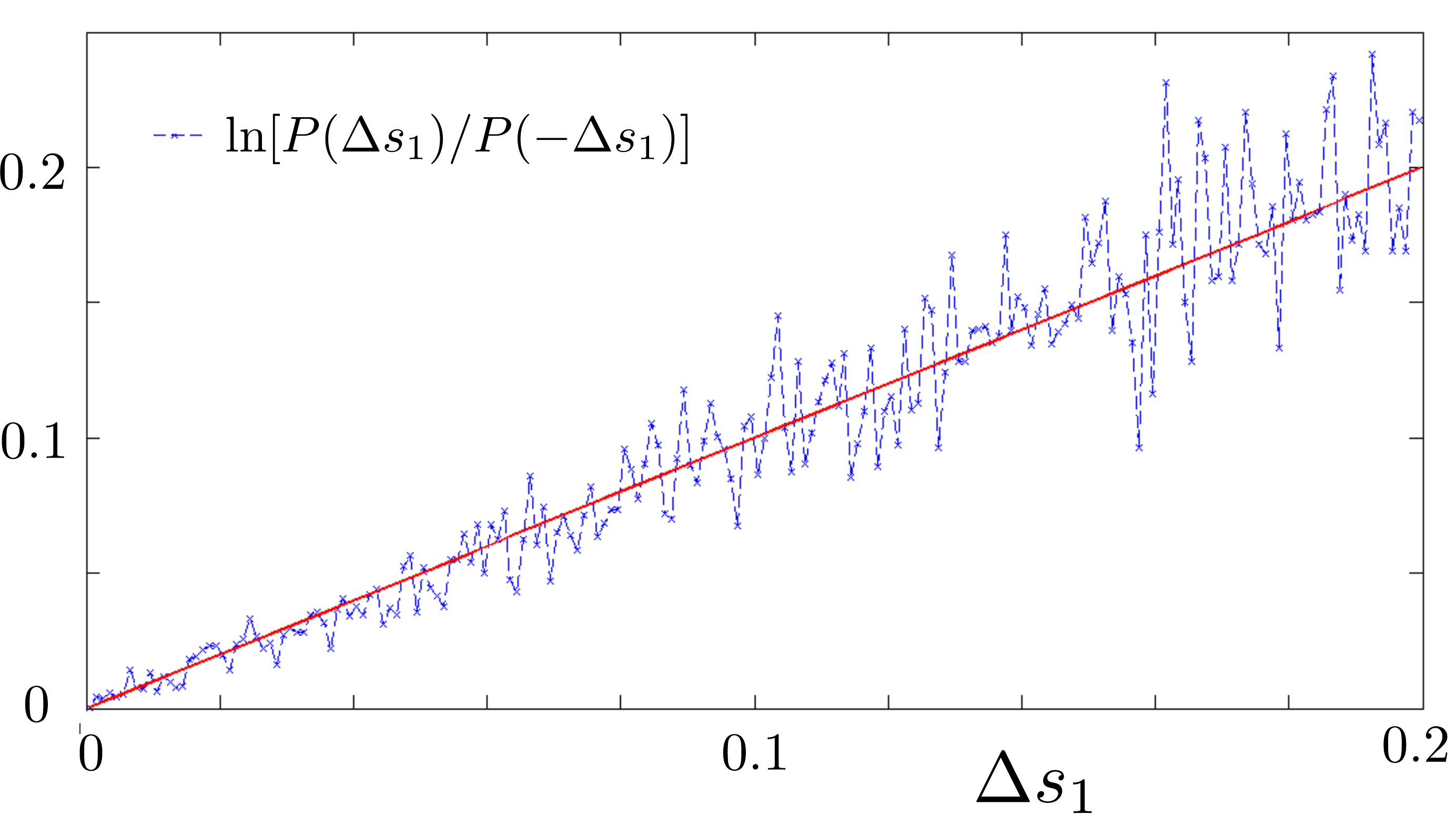} \protect\caption{As Figure \ref{fig:dftstot} but for $\Delta s_{1}$. \label{fig:dfts1}}
\end{figure}

We found that the distribution of $\Delta s_{1}$ over the same time
interval $7\pi/2\omega\le t\le11\pi/2\omega$ also appears to satisfy
a detailed fluctuation relation, as shown in Figure \ref{fig:dfts1},
while in contrast the distribution of $\Delta s_{2}$ does not possess
such a symmetry, as illustrated in Figure \ref{fig:dfts2}. In Appendix
\ref{sub:Detailed-fluctuation-relation 2} we consider conditions
for the existence of a detailed fluctuation relation for the $\Delta s_{1}$
component of entropy production in general systems with spatial and
velocity coordinates. We conclude that $P(\Delta s_{1})$ will satisfy
a detailed fluctuation relation if the friction coefficient is not
too small, such that behaviour under the chosen system dynamics and
its `adjoint' version are simply related. A detailed fluctuation relation
was observed for the total entropy production in a stationary state
of thermal transport in \cite{SpinneyFord12b}, and this can now be
interpreted as an approximate result. Furthermore, the conditions
that allow us to show that detailed fluctuation relations hold to
a certain extent for $P(\Delta s_{{\rm tot}})$ and $P(\Delta s_{1})$
do not imply a similar property for $P(\Delta s_{2})$, as shown in
Appendix \ref{sub:No-detailed-fluctuation 3}, allowing us to understand
the contrast in behaviour between Figures \ref{fig:dftstot}, \ref{fig:dfts1}
and \ref{fig:dfts2}.

\begin{figure}
\centering{} \includegraphics[width=1\columnwidth]{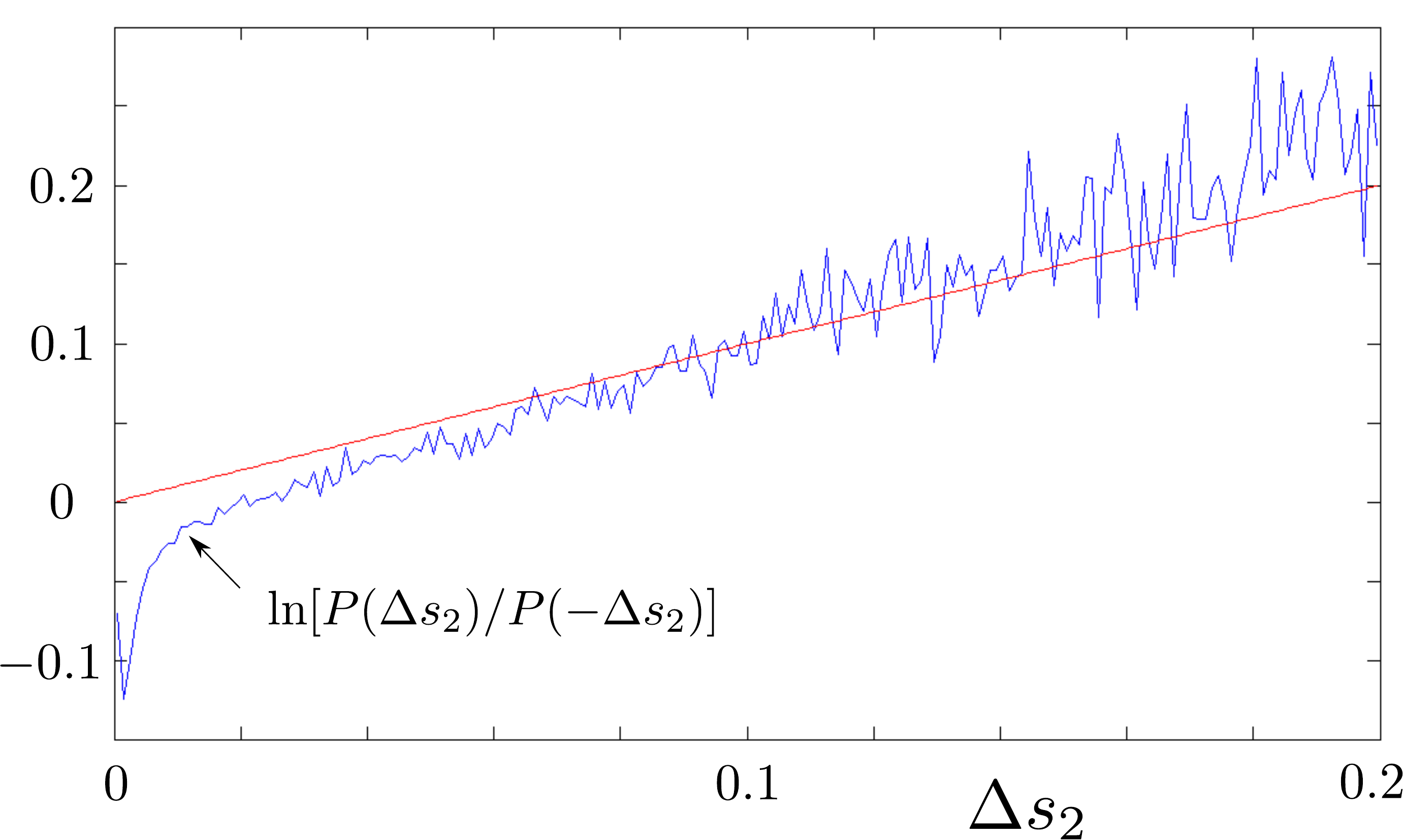} \protect\caption{Demonstration that a detailed fluctuation relation does not appear
to hold for $\Delta s_{2}$ in the interval $7\pi/2\omega\le t\le11\pi/2\omega$.
\label{fig:dfts2}}
\end{figure}

\begin{figure}
\centering{} \includegraphics[width=1\columnwidth]{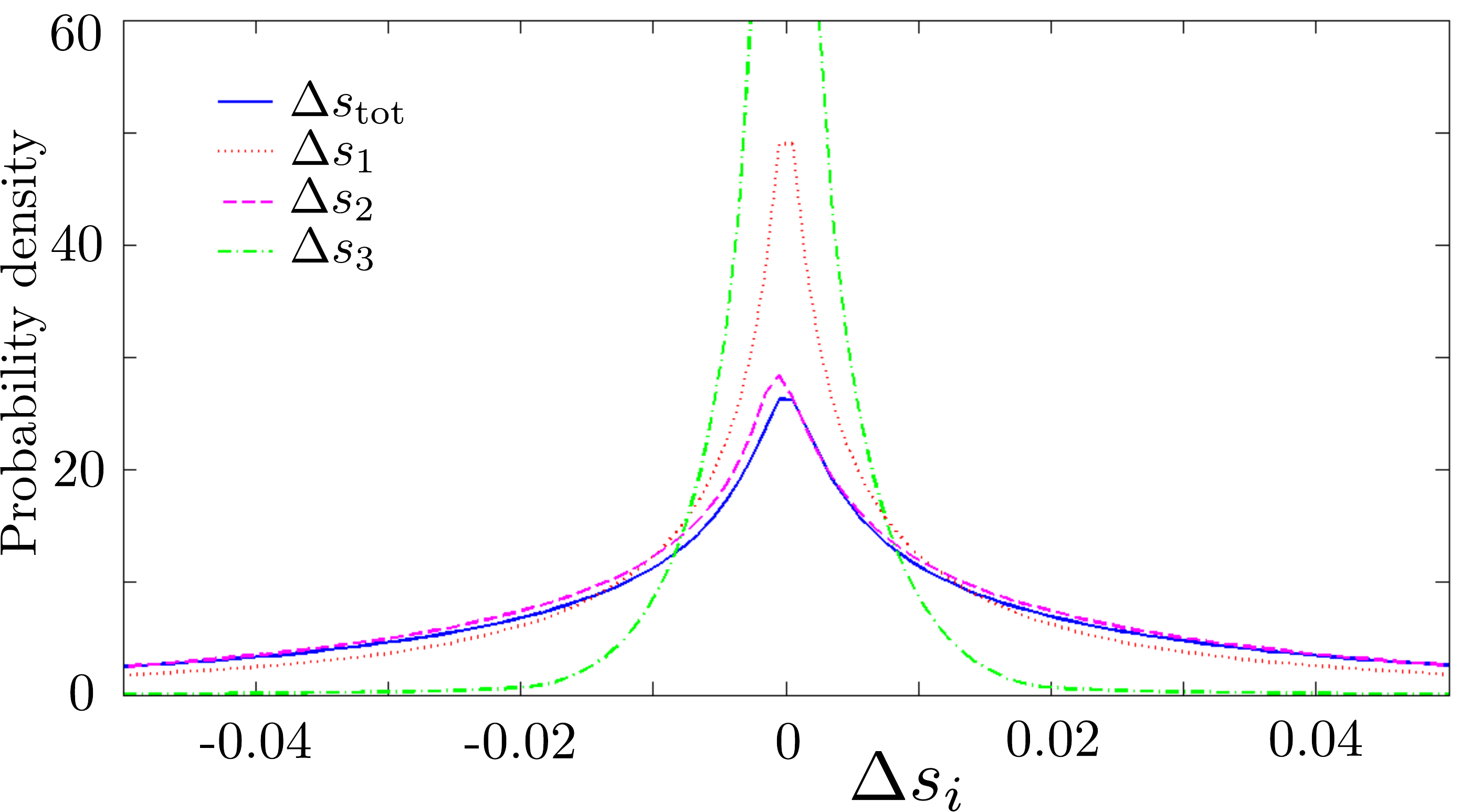} \protect\caption{Distributions of $\Delta s_{{\rm tot}}$, $\Delta s_{1}$, $\Delta s_{2}$
and $\Delta s_{3}$ generated in two cycles of temperature variation
$2\pi/\omega\le t\le6\pi/\omega$.\label{fig:Distributions}}
\end{figure}

Distributions $P(\Delta s_{i})$ of the three components of entropy
production generated over two cycles of the variation in temperature
profile, namely for the interval $2\pi/\omega\le t\le6\pi/\omega$,
together with the distribution of their sum $\Delta s_{{\rm tot}}$,
are shown in Figure \ref{fig:Distributions}. The fluctuations in
the $\Delta s_{3}$ component are the smallest, but all contributions
explore a broad range in comparison with their averages, which according
to Figures \ref{fig:s1} and \ref{fig:s2} would be of order $10^{-3}$
for $\langle\Delta s_{1}\rangle$ and $\langle\Delta s_{2}\rangle$,
together with $\langle\Delta s_{3}\rangle\approx0$.

The second law in the stochastic framework corresponds to the non-negativity
of the mean values of $\Delta s_{{\rm tot}}$, $\Delta s_{1}$ and
$\Delta s_{2}$ over distributions such as these. Although the peak
in the distribution of $\Delta s_{2}$ lies slightly to the left of
the origin, its mean is positive as required. We have clearly demonstrated
that there is considerable weight of probability for the generation
of negative values of these quantities for this small system as it
undergoes a short process. Nevertheless, such fluctuations are governed
by rules in the form of fluctuation relations of various kinds. We
have also demonstrated that the transient housekeeping component of
entropy production $\Delta s_{3}$ associated with the breakage of
the principle of detailed balance in a system evolving in full phase
space makes a significant contribution to the total stochastic entropy
production.

\section{Conclusions\label{sec:Conclusions}}

The irreversibility of a stochastic process can be quantified through
the consideration of three components of stochastic entropy production.
In order to illustrate this we have studied the behaviour of a particle
coupled to an environment characterised by a temperature that depends
on time and space. The system is complex enough to manifest all three
components of entropy production, and yet simple enough for us to
obtain an approximate expression for the time dependent probability
density function (pdf) of particle position and velocity that is required
to perform the computations. In order to solve the Kramers equation
and determine the pdf, we use a variational approach that resembles
the maximisation of an Onsager function. Such an approach has been
regarded as the use of a principle of maximisation of the rate of
thermodynamic entropy production under constraints, but there is a
certain ambiguity in the thermodynamic interpretation and we have
discussed a point of view where it might instead be regarded as a
constrained minimisation. A cautious thermodynamic interpretation
is probably necessary.

Mean relaxational or nonadiabatic entropy production $\langle\Delta s_{1}\rangle$
is driven by the time dependence of the environmental temperature,
and arises from the tendency of the system to evolve towards a state
of local thermal equilibrium with respect to the environment, which
here is frustrated by the continual environmental change. Mean housekeeping
entropy production $\langle\Delta s_{2}\rangle$ is brought about
by the spatial dependence of the environmental temperature, and is
associated with the passage of heat, on average, from hotter to cooler
parts of the environment by way of the particle. We have provided
analytic expressions, correct to first order in inverse friction coefficient,
for the evolution of the mean values of $\Delta s_{1}$ and $\Delta s_{2}$.
The mean of the third component, the transient housekeeping entropy
production $\Delta s_{3}$, is zero at the level of approximation
employed, and in order to compute a nonzero mean for this quantity
we would need to determine the system pdf to second order in inverse
friction coefficient. The $\Delta s_{3}$ component contributes to
the fluctuations in stochastic entropy production when the system
is in a stationary state characterised by a velocity asymmetric pdf,
and would be expected to have a nonzero mean when a system undergoes
relaxation: it is therefore indicative of both relaxational and housekeeping
behaviour.

We have determined the distributions of entropy production for certain
time intervals, and investigated situations where both $\Delta s_{{\rm tot}}$
and $\Delta s_{1}$ satisfy a detailed fluctuation relation. Analysis
given in Appendix \ref{sec:Detailed-fluctuation} suggests that $\Delta s_{2}$
does not have this property in the same circumstances, and that the
detailed fluctuation relations hold for the system and circumstances
under consideration as long as the friction coefficient is not too
small. Detailed fluctuation relations are a rightly celebrated centrepiece
of the thermodynamics of small systems, since they express an asymmetry
in the production and consumption of entropy, but they rely on the
validity of certain initial and final conditions for the forward and
backward processes considered, and we have illustrated this feature
for a particular underdamped system. Finally, we have computed the
distributions of the three components of entropy production, showing
that in the case studied the $\Delta s_{3}$ component has a smaller
variance than the other two.

The basis of the second law in stochastic thermodynamics is the adherence
of $\Delta s_{{\rm tot}}$, $\Delta s_{1}$ and $\Delta s_{2}$ to
integral fluctuation relations, and we have succeeded in demonstrating
the consequent monotonic increase in the mean values of these forms
of stochastic entropy production for a system processed in a way that
gives rise to nonstationary thermal transport. We have also clearly
demonstrated the existence of the third component $\Delta s_{3}$
in a situation where the stationary state of the system is asymmetric
in velocity. These observations provide further illustration of the
rich structure of stochastic thermodynamics.

\appendix

\section{Components of stochastic entropy production\label{sec:Components-of-stochastic-1}}

We summarise results that are derived in more detail in Spinney and
Ford \cite{SpinneyFord12b}, concerning the dynamics of components
of stochastic entropy production. For It$\bar{{\rm o}}$-rules stochastic
differential equations (SDEs)
\begin{equation}
d{\rm x}_{i}=A_{i}(\textbf{x},t)dt+B_{i}(\textbf{x},t)dW_{i},\label{eq:a100-1}
\end{equation}
where $\textbf{x}$ represents a set of dynamical variables $({\rm x_{1},{\rm x_{2},\cdots)}}$
such as $(x,v)$, we define
\begin{eqnarray}
A_{i}^{{\rm ir}}(\textbf{x},t) & = & \frac{1}{2}\left[A_{i}(\textbf{x},t)+\varepsilon_{i}A_{i}(\boldsymbol{\varepsilon}\textbf{x},t)\right]=\varepsilon_{i}A_{i}^{{\rm ir}}(\boldsymbol{\varepsilon}\textbf{x},t),\qquad\label{eq:a101-1}
\end{eqnarray}
and
\begin{equation}
A_{i}^{{\rm rev}}(\textbf{x},t)=\frac{1}{2}\left[A_{i}(\textbf{x},t)-\varepsilon_{i}A_{i}(\boldsymbol{\varepsilon}\textbf{x},t)\right]=-\varepsilon_{i}A_{i}^{{\rm rev}}(\boldsymbol{\varepsilon}\textbf{x},t),\quad\label{eq:a102-1}
\end{equation}
where $\varepsilon_{i}=1$ for variables ${\rm x}_{i}$ with even
parity under time reversal symmetry (for example position $x$) and
$\varepsilon_{i}=-1$ for variables with odd parity (for example velocity
$v$), and $\boldsymbol{\varepsilon}\textbf{x}$ represents $(\varepsilon_{1}{\rm x_{1},\varepsilon_{2}{\rm x_{2},\cdots)}}$.
Defining also $D_{i}(\textbf{x},t)=\frac{1}{2}B_{i}(\textbf{x},t)^{2}$,
it may be shown that the following It$\bar{{\rm o}}$-rules SDE for
the total entropy production (defined in Eq. (\ref{eq:2-1})) emerges:

\begin{eqnarray}
 &  & d\Delta s_{{\rm tot}}=-d(\ln p)+\sum_{i}\frac{A_{i}^{{\rm ir}}}{D_{i}}d{\rm x}_{i}-\frac{A_{i}^{{\rm rev}}A_{i}^{{\rm ir}}}{D_{i}}dt+\frac{\partial A_{i}^{{\rm ir}}}{\partial{\rm x}_{i}}dt\nonumber \\
 &  & -\frac{\partial A_{i}^{{\rm rev}}}{\partial{\rm x}_{i}}dt-\frac{1}{D_{i}}\frac{\partial D_{i}}{\partial{\rm x}_{i}}d{\rm x}_{i}+\frac{(A_{i}^{{\rm rev}}-A_{i}^{{\rm ir}})}{D_{i}}\frac{\partial D_{i}}{\partial{\rm x}_{i}}dt\nonumber \\
 &  & -\frac{\partial^{2}D_{i}}{\partial{\rm x}_{i}^{2}}dt+\frac{1}{D_{i}}\left(\frac{\partial D_{i}}{\partial{\rm x}_{i}}\right)^{2}dt,\quad\label{eq:a103-1}
\end{eqnarray}
where $p$ is the time dependent pdf of variables $\textbf{x}$. The
corresponding It$\bar{{\rm o}}$ SDE for the principal relaxational
entropy production is
\begin{equation}
d\Delta s_{1}=-d(\ln p)-\frac{\partial\varphi}{\partial{\rm x}_{i}}d{\rm x}_{i}-D_{i}\frac{\partial^{2}\varphi}{\partial{\rm x}_{i}^{2}}dt,\label{eq:a104-1}
\end{equation}
where $\varphi=-\ln p_{{\rm st}}$ and $p_{{\rm st}}$ is the stationary
pdf. We also have
\begin{eqnarray}
 &  & d\Delta s_{2}=\sum_{i}-\frac{A_{i}^{{\rm ir}}A_{i}^{{\rm rev}}}{D_{i}}dt+\frac{A_{i}^{{\rm ir}}}{D_{i}}d{\rm x}_{i}+\varepsilon_{i}\varphi_{i}^{\prime}(\boldsymbol{\varepsilon}\textbf{x})d{\rm x}_{i}\nonumber \\
 &  & -\frac{1}{D_{i}}\frac{\partial D_{i}}{\partial{\rm x}_{i}}d{\rm x}_{i}+\frac{1}{D_{i}}\left(\frac{\partial D_{i}}{\partial{\rm x}_{i}}\right)^{2}dt+D_{i}(\varphi_{i}^{\prime}(\boldsymbol{\varepsilon}\textbf{x}))^{2}dt\nonumber \\
 &  & -2\varepsilon_{i}\varphi_{i}^{\prime}(\boldsymbol{\varepsilon}\textbf{x})\frac{\partial D_{i}}{\partial{\rm x}_{i}}dt+\varepsilon_{i}(A_{i}^{{\rm ir}}-A_{i}^{{\rm rev}})\varphi_{i}^{\prime}(\boldsymbol{\varepsilon}\textbf{x})dt\nonumber \\
 &  & -\frac{(A_{i}^{{\rm ir}}-A_{i}^{{\rm rev}})}{D_{i}}\frac{\partial D_{i}}{\partial{\rm x}_{i}}dt,\label{eq:a105-1}
\end{eqnarray}
specifying an increment in $\Delta s_{2}$, using notation $\varphi_{i}^{\prime}(\boldsymbol{\varepsilon}\textbf{x})=\varepsilon_{i}\partial\varphi(\boldsymbol{\varepsilon}\textbf{x})/\partial{\rm x}_{i}$
and
\begin{eqnarray}
d\Delta s_{3} & = & -d\ln p_{{\rm st}}(\textbf{x})+d\ln p_{{\rm st}}(\boldsymbol{\varepsilon}\textbf{x})\nonumber \\
 & = & \sum_{i}\varphi_{i}^{\prime}(\textbf{x})\circ d{\rm x}_{i}-\varepsilon_{i}\varphi_{i}^{\prime}(\boldsymbol{\varepsilon}\textbf{x})\circ d{\rm x}_{i},\label{eq:a106-1}
\end{eqnarray}
for the third component. Stratonovich notation is used in the second
line for reasons of compactness, but a more elaborate It$\bar{{\rm o}}$-rules
version can be constructed. For the dynamics specified by Eqs. (\ref{eq:1-1})
and (\ref{eq:1a}) we have $A_{x}^{{\rm ir}}=0$, $A_{x}^{{\rm rev}}=v$,
$A_{v}^{{\rm ir}}=-\gamma v$, $A_{v}^{{\rm rev}}=F/m$, $D_{x}=0$
and $D_{v}=kT_{r}\gamma/m$ and using Eq. (\ref{eq:a103-1}) we recover
Eq. (\ref{eq:2-1}).

\section{Variational solution to the Kramers equation\label{sec:Variational-solution-to}}

We wish to obtain a solution to the approximate Kramers equation
\begin{equation}
\frac{\partial f}{\partial t}+v\frac{\partial f}{\partial x}+\left(\frac{F}{m}-\gamma\bar{v}\right)\frac{\partial f}{\partial v}=\frac{kT_{r}\gamma}{m}\frac{\partial}{\partial v}\left(f\frac{\partial(\gamma^{-1}\psi_{1})}{\partial v}\right),\label{eq:9a}
\end{equation}
that results from inserting Eq. (\ref{eq:6}) into Eq. (\ref{eq:7})
without setting $\rho=\rho_{{\rm st}}$ or $\bar{v}=0$, and using
the leading term in Eq. (\ref{eq:7a}). For simplicity of notation
we henceforth write $\psi=\gamma^{-1}\psi_{1}$. The approach involves
a variational principle employed for a similar purpose by Kohler \cite{Kohler48},
Ziman \cite{Ziman56} and Cercignani \cite{Cercignani-book2000},
and which has been discussed in a broader context by Martyushev and
Seleznev \cite{Martyushev06}. Casting Eq. (\ref{eq:9a}) in the form
$Z=\hat{L}\psi$ where $\hat{L}$ is the linear operator given by
$\hat{L}(\cdot)=(kT_{r}\gamma/m)\partial/\partial v(f\partial(\cdot)/\partial v)$,
we seek a solution by extremising the functional
\begin{equation}
I[\tilde{\psi}]=-(\tilde{\psi},Z)+\frac{1}{2}(\tilde{\psi},\hat{L}\tilde{\psi})\label{eq:9}
\end{equation}
over trial solutions $\tilde{\psi}$ that satisfy relevant constraints
$\int dvf\tilde{\psi}=0$ and $\int dvvf\tilde{\psi}=0$. The meaning
of the brackets is $(A,B)=\int dvAB$ such that
\begin{eqnarray}
(\tilde{\psi},\hat{L}\tilde{\psi}) & = & \int dv\tilde{\psi}\frac{kT_{r}\gamma}{m}\frac{\partial}{\partial v}\left(f\frac{\partial\tilde{\psi}}{\partial v}\right)\nonumber \\
 & = & -\int dv\frac{\partial\tilde{\psi}}{\partial v}\frac{kT_{r}\gamma}{m}f\frac{\partial\tilde{\psi}}{\partial v}.\label{eq:10}
\end{eqnarray}
The approach can be justified by writing $\tilde{\psi}=\psi+\eta$
in which case it may be shown that
\begin{eqnarray}
I[\tilde{\psi}] & = & I[\psi]+\frac{1}{2}(\eta,\hat{L}\eta),\label{eq:11}
\end{eqnarray}
as long as $\hat{L}$ satisfies $(\hat{L}\psi,\eta)=(\psi,\hat{L}\eta)$.
It is clear from Eq. (\ref{eq:10}) that $(\eta,\hat{L}\eta)$ for
the operator $\hat{L}$ in question is never positive, so the variational
principle may be characterised as the maximisation of $I[\tilde{\psi}]$,
which is achieved when $\tilde{\psi}=\psi$. The value of the functional
in such a circumstance is
\begin{equation}
I[\psi]=\frac{1}{2}(\psi,\hat{L}\psi)-(\psi,Z)=-\frac{1}{2}(\psi,\hat{L}\psi)\ge0.\label{eq:11a}
\end{equation}
 We clearly need to evaluate
\begin{equation}
(\tilde{\psi},Z)=\int\! dv\tilde{\psi}\frac{\partial f}{\partial t}+\int dv\tilde{\psi}v\frac{\partial f}{\partial x}+\int dv\tilde{\psi}\left(\frac{F}{m}-\gamma\bar{v}\right)\frac{\partial f}{\partial v},\label{eq:12}
\end{equation}
and for the first term we write
\begin{eqnarray}
 &  & \int dv\ \tilde{\psi}\frac{\partial f}{\partial t}=\int dv\ \tilde{\psi}\left[\frac{\partial\rho}{\partial t}\frac{f}{\rho}-\frac{f}{2T_{r}}\frac{\partial T_{r}}{\partial t}\right.\nonumber \\
 &  & \left.+f\left(\frac{m(v-\bar{v})}{kT_{r}}\frac{\partial\bar{v}}{\partial t}+\frac{m(v-\bar{v})^{2}}{2kT_{r}^{2}}\frac{\partial T_{r}}{\partial t}\right)\right]\nonumber \\
 &  & =\int dv\ \tilde{\psi}f\left(\frac{m(v-\bar{v})^{2}}{2kT_{r}^{2}}\frac{\partial T_{r}}{\partial t}\right),\label{eq:12z}
\end{eqnarray}
having used $\int dvf\tilde{\psi}=0$ and $\int dvvf\tilde{\psi}=0$.
For the second term we get
\begin{eqnarray}
 &  & \int dv\ \tilde{\psi}v\frac{\partial f}{\partial x}=\int dv\ \tilde{\psi}v\left[\frac{\partial\rho}{\partial x}\frac{f}{\rho}-\frac{f}{2T_{r}}\frac{\partial T_{r}}{\partial x}\right.\nonumber \\
 &  & \left.+f\left(\frac{m(v-\bar{v})}{kT_{r}}\frac{\partial\bar{v}}{\partial x}+\frac{m(v-\bar{v})^{2}}{2kT_{r}^{2}}\frac{\partial T_{r}}{\partial x}\right)\right]\nonumber \\
 &  & =\int dv\ \tilde{\psi}vf\left(\frac{mv}{kT_{r}}\frac{\partial\bar{v}}{\partial x}+\frac{m(v-\bar{v})^{2}}{2kT_{r}^{2}}\frac{\partial T_{r}}{\partial x}\right),\label{eq:12z-1}
\end{eqnarray}
and by similar reasoning the third term in Eq. (\ref{eq:12}) vanishes.
We therefore find that
\begin{eqnarray}
 &  & I[\tilde{\psi}]=-\frac{1}{2}\int\!\! dv\frac{\partial\tilde{\psi}}{\partial v}\frac{kT_{r}\gamma}{m}f\frac{\partial\tilde{\psi}}{\partial v}-\frac{m}{2kT_{r}^{2}}\frac{\partial T_{r}}{\partial t}\!\int\!\! dv\tilde{\psi}f(v-\bar{v}){}^{2}\nonumber \\
 &  & -\int dv\tilde{\psi}vf\left(\frac{m(v-\bar{v})}{kT_{r}}\frac{\partial\bar{v}}{\partial x}+\frac{m(v-\bar{v})^{2}}{2kT_{r}^{2}}\frac{\partial T_{r}}{\partial x}\right),\label{eq:13}
\end{eqnarray}
is the expression that has to be maximised over $\tilde{\psi}$ subject
to $\int dv\tilde{\psi}f=0$ and $\int dvv\tilde{\psi}f=0$.

The Euler-Lagrange equation that specifies the optimal $\psi$ is
\begin{eqnarray}
 &  & -\frac{kT_{r}\gamma}{m}\frac{\partial}{\partial v}f\frac{\partial\psi}{\partial v}=\lambda_{1}f+\lambda_{2}vf-\frac{m}{2kT_{r}^{2}}\frac{\partial T_{r}}{\partial t}f(v-\bar{v}){}^{2}\nonumber \\
 &  & -vf\left(\frac{m(v-\bar{v})}{kT_{r}}\frac{\partial\bar{v}}{\partial x}+\frac{m(v-\bar{v})^{2}}{2kT_{r}^{2}}\frac{\partial T_{r}}{\partial x}\right)\nonumber \\
 &  & =-\frac{kT_{r}\gamma}{m}\left(-\frac{\partial\psi}{\partial v}\frac{m(v-\bar{v})}{kT_{r}}f+f\frac{\partial^{2}\psi}{\partial v^{2}}\right),\label{eq:14}
\end{eqnarray}
where Lagrange multipliers $\lambda_{1,2}$ associated with the constraints
appear. Inserting a trial solution
\begin{equation}
\psi=\psi_{0}+a(v-\bar{v})+b(v-\bar{v})^{2}+c(v-\bar{v})^{3},\label{eq:14a}
\end{equation}
 we get
\begin{eqnarray}
 &  & \lambda_{1}+\lambda_{2}(z+\bar{v})-\frac{m}{2kT_{r}^{2}}\frac{\partial T_{r}}{\partial t}z^{2}\nonumber \\
 &  & -(z+\bar{v})\left(\frac{mz}{kT_{r}}\frac{\partial\bar{v}}{\partial x}+\frac{mz^{2}}{2kT_{r}^{2}}\frac{\partial T_{r}}{\partial x}\right)\nonumber \\
 &  & =\frac{kT_{r}\gamma}{m}\left(\left(a+2bz+3cz^{2}\right)\frac{mz}{kT_{r}}-2b-6cz\right),\qquad\label{eq:15}
\end{eqnarray}
where $z=v-\bar{v}$, and by requiring that the coefficients of $z^{2}$
and $z^{3}$ be zero together with imposing $\int dv\psi f=\int dvv\psi f=0$,
we obtain
\begin{eqnarray}
a & = & \frac{1}{2\gamma T_{r}}\frac{\partial T_{r}}{\partial x}\nonumber \\
b & = & -\frac{m}{4kT_{r}\gamma}\left(\frac{1}{T_{r}}\frac{\partial T_{r}}{\partial t}+2\frac{\partial\bar{v}}{\partial x}+\frac{\bar{v}}{T_{r}}\frac{\partial T_{r}}{\partial x}\right)\nonumber \\
c & = & -\frac{m}{6\gamma kT_{r}^{2}}\frac{\partial T_{r}}{\partial x}\nonumber \\
\psi_{0} & = & \frac{1}{4\gamma}\left(\frac{1}{T_{r}}\frac{\partial T_{r}}{\partial t}+2\frac{\partial\bar{v}}{\partial x}+\frac{\bar{v}}{T_{r}}\frac{\partial T_{r}}{\partial x}\right),\label{eq:16}
\end{eqnarray}
which specifies the approximate nonequilibrium pdf $p=f(1+\psi)$
for transient conditions. If we make the further approximations $\bar{v}=0$
and $\rho\approx\rho_{{\rm st}}$, this reduces to the solution given
in Eq. (\ref{eq:38-1}).

\section{Rate of entropy production and Onsager's principle\label{sec:Entropy-production}}

We discuss the relationship between the variational approach to solving
the Kramers equation reviewed in Appendix \ref{sec:Variational-solution-to}
and proposals for identifying a nonequilibrium stationary state based
on extremising the rate of thermodynamic entropy production. Such
a principle has been discussed many times before \cite{Dewar03,Martyushev06}.
The version that is most appropriate in the present context is the
maximisation of the Onsager function, the difference between the rate
of entropy production and a quantity denoted the dissipation function
\cite{Onsager31}. In classical nonequilibrium thermodynamics, the
rate of entropy production $\dot{S}(\{{\cal F}_{i},\tilde{J}_{i}\})$
is given by $\sum{\cal F}_{i}\tilde{J}_{i}$, in terms of currents
$\tilde{J}_{i}$ (such as the heat flux or a particle current) and
their respective driving thermodynamic forces ${\cal F}_{i}$ (gradients
in the inverse temperature field and the negative of the chemical
potential, respectively). The dissipation function is defined as $\frac{1}{2}\sum\tilde{J}_{i}L_{ij}\tilde{J}_{j}$
where the $L_{ij}$ is a matrix of coefficients. Maximisation of the
Onsager function
\begin{equation}
\Omega(\{\tilde{J}_{i}\})=\sum{\cal F}_{i}\tilde{J}_{i}-\sum\frac{1}{2}\tilde{J}_{i}L_{ij}\tilde{J}_{j},\label{eq:149-1}
\end{equation}
over currents $\tilde{J}_{i}$ for a given set of forces ${\cal F}_{i}$
produces linear relationships between the two in the form $\tilde{J}_{i}\to J_{i}=\sum R_{ij}{\cal F}_{j}$,
where $R=L^{-1}$, in agreement with the phenomenological linear response,
and hence provides a description of a nonequilibrium state. This has
been referred to as Onsager's principle and interpreted as the constrained
maximisation of the rate of production of entropy. The maximised value
of the Onsager function is $\frac{1}{2}\sum{\cal F}_{i}J_{i}$ {[}namely,
half the rate of entropy production $\dot{S}(\{{\cal F}_{i},J_{i}\})=\sum{\cal F}_{i}J_{i}${]}
if it is assumed that such linear relationships prevail, and the dissipation
function then takes the value $\frac{1}{2}\sum J_{i}L_{ij}J_{j}=\frac{1}{2}\sum{\cal F}_{i}J_{i}$.
The formalism written here in terms of summations could be taken to
apply to integrations over spatially dependent currents and forces.
However, the origin of the dissipation function and the basis of the
variational procedure are not altogether apparent.

Kohler \cite{Kohler48} and Ziman \cite{Ziman56} noticed the similarity
between the Onsager function $\Omega(\{\tilde{J}_{i}\})$ and the
variational functional $I[\tilde{\psi}]$ used in the solution of
the stochastic dynamics in Appendix \ref{sec:Variational-solution-to}
and suggested that the latter provided a microscopic dynamical underpinning
of Onsager's principle. However, the latter is clearly founded upon
a classical thermodynamic viewpoint where the primary representation
of entropy production is the product of forces and currents. In contrast,
in traditional statistical mechanics and in stochastic thermodynamics
the primary representation is given in terms of properties of the
system pdf. We can demonstrate this for the system under consideration
in this study.   By multiplying Eq. (\ref{eq:fpe}) by $-\ln p$
and integrating, we obtain after some manipulation

\begin{equation}
\frac{\partial\rho_{s}}{\partial t}+\frac{\partial j_{s}}{\partial x}=\int dv\frac{\partial\ln(1+\psi)}{\partial v}\frac{kT_{r}\gamma}{m}f\frac{\partial\psi}{\partial v}+\frac{\gamma\rho\left(T_{r}-T\right)}{T_{r}},\label{eq:19}
\end{equation}
where $\psi$ specifies the pdf that satisfies the Kramers equation,
and $\rho_{s}(x)=-\int dvp\ln p$ and $j_{s}(x)=-\int dvvp\ln p$
can be regarded as the density and current of (dimensionless) mean
system entropy, respectively. We define the local temperature of the
system $T(x,t)$ through $\int dv(v-\bar{v})^{2}p=\rho(\overline{v^{2}}-\bar{v}^{2})=kT\rho/m$.
The final term in Eq. (\ref{eq:19}) corresponds to a Clausius-style
entropy flow to the system associated with heat transfer from the
environment brought about by the difference between $T_{r}$ and $T$.
 We can therefore identify the spatial density of the rate of entropy
production from a statistical mechanical perspective, valid for transient
as well as stationary situations, as
\begin{eqnarray}
q_{s} & \approx & \int dv\frac{\partial\psi}{\partial v}\frac{kT_{r}\gamma}{m}f\frac{\partial\psi}{\partial v},\label{eq:21-1}
\end{eqnarray}
to leading order in inverse friction coefficient, which corresponds
to $\bar{v}=0$ and $\psi\ll1$. Notice that $q_{s}$ is therefore
related to the quantity

\begin{eqnarray}
-(\psi,\hat{L}\psi) & = & -\int dv\psi\frac{kT_{r}\gamma}{m}\frac{\partial}{\partial v}\left(f\frac{\partial\psi}{\partial v}\right)\approx q_{s}(\psi),\qquad\label{eq:21-2-1}
\end{eqnarray}
using the notation of Appendix \ref{sec:Variational-solution-to}.

This interpretation may be demonstrated more directly from the expression
for the mean rate of stochastic entropy production, $d\langle\Delta s_{{\rm tot}}\rangle/dt$
in Eq. (\ref{eq:stot}). We have
\begin{eqnarray}
 &  & \frac{d\langle\Delta s_{{\rm tot}}\rangle}{dt}=\int dxdv\frac{m}{pkT_{r}\gamma}\left(J_{v}^{{\rm ir}}\right)^{2}\nonumber \\
 &  & =\int dxdv\left[\frac{\partial\ln(1+\psi)}{\partial v}\frac{kT_{r}\gamma}{m}f\frac{\partial\psi}{\partial v}+\frac{m}{kT_{r}}\gamma\rho\bar{v}^{2}\right],\quad\label{eq:149}
\end{eqnarray}
which for $\bar{v}=0$ and $\psi\ll1$ reduces to the spatial integral
of $q_{s}$ given by Eq. (\ref{eq:21-1}). This formulation emphasises
that $q_{s}$ represents the spatial density of mean stochastic entropy
production for any function $\psi$ and associated pdf $p$.

These considerations provide a thermodynamic interpretation of the
$-(\tilde{\psi},\hat{L}\tilde{\psi})$ term in the functional $I[\tilde{\psi]}$
but they imply an important change in perspective. The variational
principle can also be cast as the minimisation of $-2I[\tilde{\psi}]=-(\tilde{\psi},\hat{L}\tilde{\psi})+2(\tilde{\psi},Z)=q_{s}(\tilde{\psi})+2(\tilde{\psi},Z)$,
which would then be regarded as a \emph{minimisation} of the spatial
density of the mean rate of stochastic entropy production $q_{s}(\tilde{\psi})$
with respect to the trial function $\tilde{\psi}$, under the constraint
of a fixed value of $(\tilde{\psi},Z)$. By identifying the optimal
trial function variationally, we are able to determine the solution
to the Kramers equation. As noted by Kohler, Ziman and others, such
an approach essentially provides a thermodynamic shortcut to solving
the dynamical problem.

The negative of the expression $(\tilde{\psi},Z)$, when restricted
to a stationary state with $\partial T_{r}/\partial t=0$ and $\bar{v}=0$,
is given by
\begin{eqnarray}
-(\tilde{\psi},Z) & \approx & -\frac{\partial T_{r}}{\partial x}\int dv\tilde{\psi}f\frac{mv^{3}}{2kT_{r}^{2}}\nonumber \\
 & = & \frac{\partial(T_{r}^{-1})}{\partial x}\int dv\left[f+\tilde{\psi}f\right]\frac{mv^{3}}{2k},\label{eq:150}
\end{eqnarray}
which is the product of the gradient of inverse temperature and the
particle kinetic energy flux (divided by $k$) at position $x$, therefore
resembling the spatial density of an entropy production rate in classical
thermodynamics, for a trial $\tilde{\psi}$. However, in the context
of stochastic thermodynamics $-(\tilde{\psi},Z)$ is not to be primarily
identified as the entropy production, whereas $q_{s}$ most definitely
can be so interpreted.

We have followed Kohler and Ziman in regarding the variational procedure
used in Appendix \ref{sec:Variational-solution-to} as a principle
founded in dynamics (i.e. $\tilde{\psi}$ is selected in order to
satisfy the Kramers equation) but which is capable of a thermodynamic
interpretation. However, the apparent thermodynamic principle in operation
is not quite as clear as has been suggested.  The semantic subtlety
is whether, following Onsager, we must maximise classical entropy
production $\sum{\cal F}_{i}\tilde{J}_{i}$, subject to a fixed dissipation
function $\frac{1}{2}\sum\tilde{J}_{i}L_{ij}\tilde{J}_{j}$, over
currents $\tilde{J_{i}}$ for a given set of forces ${\cal F}_{i}$,
or minimise a stochastic thermodynamic representation of the spatial
density of entropy production, in this case $q_{s}(\tilde{\psi})$,
with respect to a function $\tilde{\psi}$ subject to a fixed $(\tilde{\psi},Z)$.
The fact that both interpretations can be maintained, depending on
whether we use a classical or a stochastic framework of entropy production,
suggests that taking a particular thermodynamic viewpoint of the procedure
should be treated with caution.

\section{Detailed fluctuation relations\label{sec:Detailed-fluctuation}}

\subsection{Detailed fluctuation relation for $\Delta s_{{\rm tot}}$\label{sub:Detailed-fluctuation-relation 1}}

The total entropy production associated with a trajectory takes the
form
\begin{equation}
\Delta s_{{\rm tot}}^{{\rm F}}[\boldsymbol{X}]=\ln\left[\frac{p_{{\rm start}}^{{\rm F}}[X(0)]{\cal P}^{{\rm F}}[X(\tau)|X(0)]}{p_{{\rm end}}^{{\rm F}}[X(\tau)]{\cal P}^{{\rm R}}[X^{\dagger}(\tau)|X^{\dagger}(0)]}\right],\label{eq:200-2}
\end{equation}
where $\boldsymbol{X}$ corresponds to $\vec{\boldsymbol{x}},\vec{\boldsymbol{v}}$
and $X(t)$ represents the coordinates $x(t),v(t)$ taken by the system
under what we shall call a forward protocol of driving, while $\boldsymbol{X}^{\dagger}$
corresponds to $\vec{\boldsymbol{x}}^{\dagger},\vec{\boldsymbol{v}}^{\dagger}$
with $X^{\dagger}(t)$ representing $x^{\dagger}(t),v^{\dagger}(t)$
where $x^{\dagger}(t)=x(\tau-t)$ and $v^{\dagger}(t)=-v(\tau-t)$.
The superscript R indicates driving according to the reverse of the
protocol that operates in the forward process, which is labelled F.
The above expression is compatible with Eq. (\ref{eq:2-1}) with ${\rm P^{{\rm F}}}[\vec{\boldsymbol{x}},\vec{\boldsymbol{v}}]$
here explicitly written as $p_{{\rm start}}^{{\rm F}}[X(0)]{\cal P}^{{\rm F}}[X(\tau)|X(0)]$
in terms of a pdf of initial coordinates $p_{{\rm start}}^{{\rm F}}$
and a conditional probability density ${\cal P}^{{\rm F}}$. Note
that the expression takes the form of a ratio of probabilities of
a trajectory and a (nominal) reverse or antitrajectory, the initial
pdf of which is $p_{{\rm end}}^{{\rm F}}[X(\tau)]$, the pdf of coordinates
at the end of the forward trajectory. There is an implied inversion
\cite{Ford15c} of the velocity coordinate such that $X(\tau)\to X^{\dagger}(0)$
before the continuation with the reverse trajectory $X^{\dagger}(0)\to X^{\dagger}(\tau)$.
The pdf of initial coordinates for the reverse trajectory is therefore
determined by the forward process.

The distribution of entropy production for the forward process can
be written as
\begin{eqnarray}
 &  & P^{{\rm F}}(\Delta s_{{\rm tot}}^{{\rm F}}=A)\label{eq:202-2}\\
 &  & =\int d\boldsymbol{X}\; p_{{\rm start}}^{{\rm F}}[X(0)]{\cal P}^{{\rm F}}[X(\tau)|X(0)]\delta(A-\Delta s_{{\rm tot}}^{{\rm F}}[\boldsymbol{X}]),\nonumber
\end{eqnarray}
and this depends on the form taken by the initial pdf, and the nature
of the prevailing dynamics.

We next consider the entropy production for a trajectory generated
in a process starting from a pdf $p_{{\rm start}}^{{\rm R}}$ and
driven by a reverse protocol. This is
\begin{equation}
\Delta s_{{\rm tot}}^{{\rm R}}[\bar{\boldsymbol{X}}]=\ln\left[\frac{p_{{\rm start}}^{{\rm R}}[\bar{X}(0)]{\cal P}^{{\rm R}}[\bar{X}(\tau)|\bar{X}(0)]}{p_{{\rm end}}^{{\rm R}}[\bar{X}(\tau)]{\cal P}^{{\rm F}}[\bar{X}^{\dagger}(\tau)|\bar{X}^{\dagger}(0)]}\right].\label{eq:202a}
\end{equation}
  The trajectory shown here is general but it will prove fruitful
to write $\bar{\boldsymbol{X}}$ as $\boldsymbol{X}^{\dagger}$ and
therefore related to $\boldsymbol{X}$, i.e. $\bar{X}(0)$ represents
$x(\tau),-v(\tau)$; $\bar{X}(\tau)$ is $x(0),-v(0)$; $\bar{X}^{\dagger}(0)$
represents $x(0),v(0)$ or $X(0)$; and $\bar{X}^{\dagger}(\tau)$
is $x(\tau),v(\tau)$ or $X(\tau)$. Clearly this specification of
$\bar{\boldsymbol{X}}$ satisfies the dynamics under reverse driving.
 We have
\begin{eqnarray}
\Delta s_{{\rm tot}}^{{\rm R}}[\boldsymbol{X}^{\dagger}] & = & \ln\left[\frac{p_{{\rm start}}^{{\rm R}}[X^{\dagger}(0)]{\cal P}^{{\rm R}}[X^{\dagger}(\tau)|X^{\dagger}(0)]}{p_{{\rm end}}^{{\rm R}}[X^{\dagger}(\tau)]{\cal P}^{{\rm F}}[X(\tau)|X(0)]}\right],\quad\label{eq:202b}
\end{eqnarray}
 and we compute the distribution of total entropy production in
this process:
\begin{eqnarray}
 &  & P^{{\rm R}}(\Delta s_{{\rm tot}}^{{\rm R}}=A)\label{eq:206-1-1}\\
 &  & \!=\!\int\! d\boldsymbol{X}^{\dagger}p_{{\rm start}}^{{\rm R}}[X^{\dagger}(0)]{\cal P}^{{\rm R}}[X^{\dagger}(\tau)|X^{\dagger}(0)]\delta(A-\Delta s_{{\rm tot}}^{{\rm R}}[\boldsymbol{X}^{\dagger}]).\nonumber
\end{eqnarray}
Writing Eq. (\ref{eq:202b}) in the form
\begin{eqnarray}
 &  & p_{{\rm start}}^{{\rm R}}[X^{\dagger}(0)]{\cal P}^{{\rm R}}[X^{\dagger}(\tau)|X^{\dagger}(0)]\nonumber \\
 &  & =p_{{\rm end}}^{{\rm R}}[X^{{\rm \dagger}}(\tau)]{\cal P}^{{\rm F}}[X(\tau)|X(0)]{\rm e}^{\Delta s_{{\rm tot}}^{{\rm R}}[\boldsymbol{X}^{\dagger}]},\label{eq:207-2}
\end{eqnarray}
 and noting that $d\boldsymbol{X}=d\boldsymbol{X}^{\dagger}$, we
find that
\begin{eqnarray}
 &  & P^{{\rm R}}(\Delta s_{{\rm tot}}^{{\rm R}}=A)=\int d\boldsymbol{X}^{\dagger}\; p_{{\rm end}}^{{\rm R}}[X^{{\rm \dagger}}(\tau)]{\cal P}^{{\rm F}}[X(\tau)|X(0)]\nonumber \\
 &  & \qquad\times{\rm e}^{\Delta s_{{\rm tot}}^{{\rm R}}[\boldsymbol{X}^{\dagger}]}\delta(A-\Delta s_{{\rm tot}}^{{\rm R}}[\boldsymbol{X}^{\dagger}])\label{eq:207a-2a}\\
 &  & \!\!={\rm e}^{A}\int d\boldsymbol{X}p_{{\rm end}}^{{\rm R}}[X^{{\rm \dagger}}(\tau)]{\cal P}^{{\rm F}}[X(\tau)|X(0)]\delta(A-\Delta s_{{\rm tot}}^{{\rm R}}[\boldsymbol{X}^{\dagger}]).\nonumber
\end{eqnarray}

Now, if it can be arranged that $p_{{\rm end}}^{{\rm R}}[X^{\dagger}(\tau)]=p_{{\rm start}}^{{\rm F}}[X(0)]$
or $p_{{\rm end}}^{{\rm R}}[x,-v]=p_{{\rm start}}^{{\rm F}}[x,v]$;
and $p_{{\rm start}}^{{\rm R}}[X^{{\rm \dagger}}(0)]=p_{{\rm end}}^{{\rm F}}[X(\tau)]$
or $p_{{\rm start}}^{{\rm R}}[x,-v]=p_{{\rm end}}^{{\rm F}}[x,v]$,
then it would follow that
\begin{eqnarray}
\Delta s_{{\rm tot}}^{{\rm R}}[\boldsymbol{X}^{\dagger}] & = & \ln\left[\frac{p_{{\rm end}}^{{\rm F}}[X(\tau)]{\cal P}^{{\rm R}}[X^{\dagger}(\tau)|X^{\dagger}(0)]}{p_{{\rm start}}^{{\rm F}}[X(0)]{\cal P}^{{\rm F}}[X(\tau)|X(0)]}\right]\nonumber \\
 & = & -\Delta s_{{\rm tot}}^{{\rm F}}[\boldsymbol{X}],\label{eq:207b}
\end{eqnarray}
in which case we would obtain the detailed fluctuation relation
\begin{eqnarray}
 &  & P^{{\rm R}}(\Delta s_{{\rm tot}}^{{\rm R}}=A)=\nonumber \\
 &  & {\rm e}^{A}\!\int d\boldsymbol{X}\, p_{{\rm start}}^{{\rm F}}[X(0)]{\cal P}^{{\rm F}}[X(\tau)|X(0)]\delta(A+\Delta s_{{\rm tot}}^{{\rm F}}[\boldsymbol{X}])\nonumber \\
 &  & ={\rm e}^{A}\, P^{{\rm F}}(\Delta s_{{\rm tot}}^{{\rm F}}=-A).\label{eq:207c}
\end{eqnarray}
  It is necessary to state clearly what this means. It relates the
pdf of total entropy production in a forward process to the pdf of
total entropy production when the system is driven by a reverse protocol
instead, with the condition that the initial pdf for the reverse process
is a velocity inverted version of the final pdf in the forward process,
and similarly the initial pdf for the forward process is a velocity
inverted version of the final pdf from the reverse process. We have
essentially followed \cite{Harris07} and \cite{Shargel10} in this
derivation.

A single system can be subjected to a forward and reverse process
sequentially such that $p_{{\rm start}}^{{\rm R}}[x,v]=p_{{\rm end}}^{{\rm F}}[x,v]$
but the condition for the detailed fluctuation relation to hold would
then require the pdf at the end of the forward sequence to be velocity
symmetric. For a system driven by a repeated sequence of forward and
reverse processes, $p_{{\rm start}}^{{\rm F}}[x,v]=p_{{\rm end}}^{{\rm R}}[x,v]$
would apply as well, implying velocity symmetry in the pdf at the
end of the reverse sequence. Furthermore, if the forward and reverse
protocols are identical, which requires each to be symmetric about
the midpoint in the interval, the condition for the validity of the
detailed fluctuation relation requires the initial and final pdfs
to be the same, in which case Eq. (\ref{eq:207c}) would reduce to
\begin{equation}
P(\Delta s_{{\rm tot}})={\rm e}^{\Delta s_{{\rm tot}}}\, P(-\Delta s_{{\rm tot}}),\label{eq:213-2}
\end{equation}
for such an interval, which is the detailed fluctuation relation that
is tested in Section \ref{sec:Distributions-of-stochastic}. The required
velocity symmetry of the pdf at the beginning and end of the interval
does not hold in general, and indeed is not satisfied by $p(x,v,t)$
in Eq. (\ref{eq:38-1}) for the particular system we have studied,
but for large $\gamma$ the asymmetry is small and in such circumstances
Eq. (\ref{eq:213-2}) will apply to an approximate extent.

\subsection{Detailed fluctuation relation for $\Delta s_{1}$\label{sub:Detailed-fluctuation-relation 2}}

We can perform a similar analysis to find conditions for which the
relaxational entropy production satisfies a detailed fluctuation relation.
We start with the definition \cite{Jarpathintegral,SpinneyFord12b}
\begin{equation}
\Delta s_{1}^{{\rm F}}[\boldsymbol{X}]=\ln\left[\frac{p_{{\rm start}}^{{\rm F}}[X(0)]{\cal P}^{{\rm F}}[X(\tau)|X(0)]}{p_{{\rm end}}^{{\rm F}}[X(\tau)]{\cal P}_{{\rm ad}}^{{\rm R}}[X^{{\rm R}}(\tau)|X^{{\rm R}}(0)]}\right],\label{eq:200-1-1}
\end{equation}
where $\boldsymbol{X}^{{\rm R}}$ corresponds to $\vec{\boldsymbol{x}}^{{\rm R}},\vec{\boldsymbol{v}}^{{\rm R}}$
with $X^{{\rm R}}(t)$ representing $x^{{\rm R}}(t),v^{{\rm R}}(t)$
where $x^{{\rm R}}(t)=x(\tau-t)$ and $v^{{\rm R}}(t)=v(\tau-t)$.
Introducing further notation, we have $X^{{\rm R}}(t)=\hat{T}X^{{\rm \dagger}}(t)$
where $\hat{T}$ is an operator that reverses the sign of velocity
coordinates such that $\hat{T}x=x$ and $\hat{T}v=-v$. The subscript
ad indicates that the trajectory $\boldsymbol{X}^{{\rm R}}$ is generated
according to \emph{adjoint} dynamics, to be discussed shortly. The
distribution of relaxational entropy production for the forward process
is
\begin{eqnarray}
 &  & P^{{\rm F}}(\Delta s_{1}=A)\label{eq:202-1-1}\\
 &  & =\int d\boldsymbol{X}\; p_{{\rm start}}^{{\rm F}}[X(0)]{\cal P}^{{\rm F}}[X(\tau)|X(0)]\delta(A-\Delta s_{1}[\boldsymbol{X}]).\nonumber
\end{eqnarray}
Next we consider relaxational entropy production for a system with
a starting pdf $p_{{\rm start}}^{{\rm R}}$ and driven by a reverse
protocol. This is
\begin{equation}
\Delta s_{1}^{{\rm R}}[\bar{\boldsymbol{X}}]=\ln\left[\frac{p_{{\rm start}}^{{\rm R}}[\bar{X}(0)]{\cal P}^{{\rm R}}[\bar{X}(\tau)|\bar{X}(0)]}{p_{{\rm end}}^{{\rm R}}[\bar{X}(\tau)]{\cal P_{{\rm ad}}^{{\rm F}}}[\bar{X}^{{\rm R}}(\tau)|\bar{X}^{{\rm R}}(0)]}\right].\label{eq:202-1a}
\end{equation}
Once again we represent $\bar{\boldsymbol{X}}$ as $\boldsymbol{X}^{\dagger}$
i.e. $\bar{X}(0)$ represents $x(\tau),-v(\tau)$, or $\hat{T}X(\tau)$;
$\bar{X}(\tau)$ is $x(0),-v(0)$, or $\hat{T}X(0)$; $\bar{X}^{{\rm R}}(0)$
represents $x(0),-v(0)$, and also $\hat{T}X(0)$; and $\bar{X}^{{\rm R}}(\tau)$
is $x(\tau),-v(\tau)$ or $\hat{T}X(\tau)$. We write
\begin{eqnarray}
\Delta s_{{\rm 1}}^{{\rm R}}[\boldsymbol{X}^{{\rm \dagger}}] & = & \ln\left[\frac{p_{{\rm start}}^{{\rm R}}[\hat{T}X(\tau)]{\cal P}^{{\rm R}}[\hat{T}X(0)|\hat{T}X(\tau)]}{p_{{\rm end}}^{{\rm R}}[\hat{T}X(0)]{\cal P_{{\rm ad}}^{{\rm F}}}[\hat{T}X(\tau)|\hat{T}X(0)]}\right].\qquad\quad\label{eq:202-1b}
\end{eqnarray}
Now, if it can be arranged that $p_{{\rm start}}^{{\rm R}}[\hat{T}X(\tau)]=p_{{\rm end}}^{{\rm F}}[X(\tau)]$
or $p_{{\rm start}}^{{\rm R}}[x,-v]=p_{{\rm end}}^{{\rm F}}[x,v]$;
and $p_{{\rm end}}^{{\rm R}}[\hat{T}X(0)]=p_{{\rm start}}^{{\rm F}}[X(0)]$
or $p_{{\rm end}}^{{\rm R}}[x,-v]=p_{{\rm start}}^{{\rm F}}[x,v]$;
together with ${\cal P_{{\rm ad}}^{{\rm F}}}[\hat{T}X(\tau)|\hat{T}X(0)]={\cal P^{{\rm F}}}[X(\tau)|X(0)]$
and ${\cal P}^{{\rm R}}[\hat{T}X(0)|\hat{T}X(\tau)]={\cal P_{{\rm ad}}^{{\rm R}}}[X(0)|X(\tau)]$,
certainly a demanding set of conditions, then we would be able to
write, using $X^{{\rm R}}(\tau)=X(0)$ and $X^{{\rm R}}(0)=X(\tau)$,
\begin{eqnarray}
\Delta s_{{\rm 1}}^{{\rm R}}[\boldsymbol{X}^{\dagger}] & = & \ln\left[\frac{p_{{\rm end}}^{{\rm F}}[X(\tau)]{\cal P_{{\rm ad}}^{{\rm R}}}[X^{{\rm R}}(\tau)|X^{{\rm R}}(0)]}{p_{{\rm start}}^{{\rm F}}[X(0)]{\cal P^{{\rm F}}}[X(\tau)|X(0)]}\right]\nonumber \\
 & = & -\Delta s_{{\rm 1}}^{{\rm F}}[\boldsymbol{X}].\label{eq:202-1c}
\end{eqnarray}
We compute the distribution of relaxational entropy production in
the reverse process in these circumstances:
\begin{eqnarray}
 &  & P^{{\rm R}}(\Delta s_{{\rm 1}}^{{\rm R}}=A)\label{eq:206a}\\
 &  & =\!\int d\boldsymbol{X}^{{\rm \dagger}}p_{{\rm start}}^{{\rm R}}[X^{{\rm \dagger}}(0)]{\cal P}^{{\rm R}}[X^{{\rm \dagger}}(\tau)|X^{{\rm \dagger}}(0)]\delta(A-\Delta s_{{\rm 1}}^{{\rm R}}[\boldsymbol{X}^{{\rm \dagger}}]),\nonumber
\end{eqnarray}
and writing
\begin{eqnarray}
 &  & p_{{\rm start}}^{{\rm R}}[X^{{\rm \dagger}}(0)]{\cal P^{{\rm R}}}[X^{{\rm \dagger}}(\tau)|X^{{\rm \dagger}}(0)]\nonumber \\
 &  & =p_{{\rm start}}^{{\rm R}}[\hat{T}X(\tau)]{\cal P^{{\rm R}}}[\hat{T}X(0)|\hat{T}X(\tau)]\nonumber \\
 &  & =p_{{\rm end}}^{{\rm F}}[X(\tau)]{\cal P_{{\rm ad}}^{{\rm R}}}[X^{{\rm R}}(\tau)|X^{{\rm R}}(0)]\nonumber \\
 &  & =p_{{\rm start}}^{{\rm F}}[X(0)]{\cal P^{{\rm F}}}[X(\tau)|X(0)]{\rm e}^{-\Delta s_{{\rm 1}}^{{\rm F}}[\boldsymbol{X}]},\label{eq:207-1-1}
\end{eqnarray}
which employs several of our assumptions, and also using $\Delta s_{{\rm 1}}^{{\rm R}}[\boldsymbol{X}^{{\rm \dagger}}]=-\Delta s_{{\rm 1}}^{{\rm F}}[\boldsymbol{X}]$,
we find that
\begin{eqnarray}
 &  & P^{{\rm R}}(\Delta s_{{\rm 1}}^{{\rm R}}=A)=\int d\boldsymbol{X}\; p_{{\rm start}}^{{\rm F}}[X(0)]{\cal P^{{\rm F}}}[X(\tau)|X(0)]\nonumber \\
 &  & \qquad\times{\rm e}^{-\Delta s_{{\rm 1}}^{{\rm F}}[\boldsymbol{X}]}\delta(A+\Delta s_{{\rm 1}}^{{\rm F}}[\boldsymbol{X}])\label{eq:207a-1-1a}\\
 &  & \!={\rm e}^{A}\!\int d\boldsymbol{X}p_{{\rm start}}^{{\rm F}}[X(0)]{\cal P}^{{\rm F}}[X(\tau)|X(0)]\delta(A+\Delta s_{{\rm 1}}^{{\rm F}}[\boldsymbol{X}]),\nonumber
\end{eqnarray}
and hence obtain the detailed fluctuation relation
\begin{equation}
P^{{\rm R}}(\Delta s_{{\rm 1}}^{{\rm R}}=A)={\rm e}^{A}\, P^{{\rm F}}(\Delta s_{{\rm 1}}^{{\rm F}}=-A).\label{transient208-1-1}
\end{equation}
This relates the pdf of relaxational entropy production in a backward
process starting from $p_{{\rm start}}^{{\rm R}}$ to the pdf of relaxational
entropy production when the system is driven by a forward protocol
starting from $p_{{\rm start}}^{{\rm F}}$, subject to the assumptions
made about the relationships between initial and final pdfs and the
normal and adjoint dynamics.

For this fluctuation relation to apply to the case of a single system
driven by a sequence of forward and backward protocols, we require
$p_{{\rm start}}^{{\rm R}}$ to equal $p_{{\rm end}}^{{\rm F}}$ and
$p_{{\rm start}}^{{\rm F}}$ to equal $p_{{\rm end}}^{{\rm R}}$.
Bearing in mind the requirements placed on the pdfs, both $p_{{\rm start}}^{{\rm F}}$
and $p_{{\rm start}}^{{\rm R}}$ should be velocity symmetric, as
we found when considering the total entropy production, and this will
hold to an approximation for large $\gamma$ for the system we consider
here.

The conditions ${\cal P_{{\rm ad}}^{{\rm F}}}[\hat{T}X(\tau)|\hat{T}X(0)]={\cal P^{{\rm F}}}[X(\tau)|X(0)]$
and ${\cal P}^{{\rm R}}[\hat{T}X(0)|\hat{T}X(\tau)]={\cal P_{{\rm ad}}^{{\rm R}}}[X(0)|X(\tau)]$,
or in more compact form ${\cal P_{{\rm ad}}}[\hat{T}X^{\prime}|\hat{T}X]={\cal P}[X^{\prime}|X]$,
can also be justified for large $\gamma$ in this system. Adjoint
dynamics are constructed from normal dynamics in order to preserve
a stationary pdf $p_{{\rm st}}$ but to reverse the probability current
\cite{Jarpathintegral,Harris07,adiabaticnonadiabatic0}. In general,
if normal dynamics correspond to the SDEs $d{\rm x}_{i}=A_{i}dt+B_{i}dW_{i}$
for dynamical variables ${\rm x}_{i}$ then the adjoint dynamics are
described by $d{\rm x}_{i}=A_{i}^{{\rm ad}}dt+B_{i}dW_{i}$ \cite{SpinneyFord12b}
with
\begin{equation}
A_{i}^{{\rm ad}}=-A_{i}+2\frac{\partial D_{i}}{\partial{\rm x}_{i}}+2D_{i}\frac{\partial\ln p_{{\rm st}}}{\partial{\rm x}_{i}},\label{eq:301}
\end{equation}
where $D_{i}=\frac{1}{2}B_{i}^{2}$. We note that for normal dynamics
given by $dx=vdt$ and
\begin{equation}
dv=-\gamma vdt+\frac{F}{m}dt+\left(\frac{2kT_{r}\gamma}{m}\right)^{1/2}dW,\label{eq:302}
\end{equation}
corresponding to $A_{x}=v$, $D_{x}=0$, $A_{v}=-\gamma v+F/m$, $D_{v}=kT_{r}\gamma/m$,
and $p_{{\rm st}}\propto[1+O(\gamma^{-1})]\exp(-mv^{2}/2kT_{r})$
according to Eq. (\ref{eq:38-2}), then $A_{x}^{{\rm ad}}=-v$ and
\begin{eqnarray}
\!\! A_{v}^{{\rm ad}} & \!= & \gamma v-\frac{F}{m}-\frac{2kT_{r}\gamma}{m}\frac{mv}{kT_{r}}\!+\! O(\gamma^{-1})\approx\!-\gamma v-\frac{F}{m}.\qquad\,\,\label{eq:303-1}
\end{eqnarray}
The SDEs for the adjoint dynamics are therefore $dx=-vdt$ and
\begin{equation}
dv\approx-\gamma vdt-\frac{F}{m}dt+\left(\frac{2kT_{r}\gamma}{m}\right)^{1/2}dW,\label{eq:304}
\end{equation}
or $dx=Vdt$ together with
\begin{equation}
dV\approx-\gamma Vdt+\frac{F}{m}dt-\left(\frac{2kT_{r}\gamma}{m}\right)^{1/2}dW,\label{eq:305}
\end{equation}
where $V=-v$. Comparing with the SDEs for the normal dynamics, it
is clear that to order $\gamma^{-1}$ the velocity inverted coordinates
evolve under adjoint dynamics in the same way that the original coordinates
evolve under normal dynamics (the change in sign of the noise term
is irrelevant). This is precisely the meaning of the condition ${\cal P_{{\rm ad}}}[\hat{T}X^{\prime}|\hat{T}X]={\cal P}[X^{\prime}|X]$
and we conclude that the detailed fluctuation relation (\ref{transient208-1-1})
should hold for large $\gamma$. It follows that if the forward and
reverse protocols are identical then this relation becomes
\begin{equation}
P(\Delta s_{{\rm 1}})={\rm e}^{\Delta s_{1}}\, P(-\Delta s_{{\rm 1}}),\label{eq:213-1-1}
\end{equation}
which is the detailed fluctuation relation that is investigated in
Section \ref{sec:Distributions-of-stochastic}.

\bigskip{}

\subsection{No detailed fluctuation relation for $\Delta s_{2}$ \label{sub:No-detailed-fluctuation 3}}

The principal component of housekeeping entropy production \cite{SpinneyFord12a,SpinneyFord12b}
takes the form
\begin{equation}
\Delta s_{2}^{{\rm F}}[\boldsymbol{X}]=\ln\left[\frac{{\cal P}^{{\rm F}}[X(\tau)|X(0)]}{{\cal P}_{{\rm ad}}^{{\rm F}}[X^{{\rm T}}(\tau)|X^{{\rm T}}(0)]}\right],\label{eq:401}
\end{equation}
where $\boldsymbol{X}^{{\rm T}}$ corresponds to $\vec{\boldsymbol{x}}^{{\rm T}},\vec{\boldsymbol{v}}^{{\rm T}}$
with $X^{{\rm T}}(t)$ representing $x^{{\rm T}}(t),v^{{\rm T}}(t)$
where $x^{{\rm T}}(t)=x(t)$ and $v^{{\rm T}}(t)=-v(t)$. The subscript
ad once again indicates that the trajectory in question is to be generated
according to adjoint dynamics.

The denominator in Eq. (\ref{eq:401}) may be written ${\cal P}_{{\rm ad}}^{{\rm F}}[\hat{T}X(\tau)|\hat{T}X(0)]$
and for small $\gamma^{-1}$ we saw previously that for our system
this is approximately equal to ${\cal P}^{{\rm F}}[X(\tau)|X(0)]$,
which appears in the numerator. The approximations that support the
validity of detailed fluctuation relations for $\Delta s_{{\rm tot}}$
and $\Delta s_{1}$ therefore suggest that $\Delta s_{2}$ vanishes,
giving us reason not to expect a detailed fluctuation relation for
$\Delta s_{2}$ and to understand the contrast between Figures \ref{fig:dftstot},
\ref{fig:dfts1} and \ref{fig:dfts2}.


%

\end{document}